\documentstyle[12pt,epsfig,psfrag,amsmath,pifont,amssymb,cite]{article}
\numberwithin{equation}{section}
\numberwithin{table}{section}
\def\ga{\mathrel{\raise.3ex\hbox{$>$\kern-.75em\lower1ex\hbox{$\sim$}}}}
\def\la{\mathrel{\raise.3ex\hbox{$<$\kern-.75em\lower1ex\hbox{$\sim$}}}}

\def\I_M{{I_{\scriptscriptstyle M\times M}}}

\def\be{\begin{equation}}
\def\ee{\end{equation}}
\def\bea{\begin{eqnarray}}
\def\eea{\end{eqnarray}}

\newcommand{\beqal}{\begin{eqnarray}\label}
\newcommand{\beqa}{\begin{eqnarray}}
\newcommand{\eeqa}{\end{eqnarray}}
\newcommand {\br}{\bar r}
\newcommand {\bq}{\bar{q}}
\newcommand {\bT}{\bar{T}}
\newcommand {\bmu}{\bar{\mu}}
\newcommand {\bF}{\bar{\cal{F}}}

\begin{document}

\begin{titlepage}
\begin{center}
\vskip .2in

{\Large \bf Notes on R-charged black holes near criticality and gauge theory}
\vskip .5in

{\bf Sachin Jain\footnote{e-mail: sachjain@iopb.res.in}, \bf Sudipta 
Mukherji\footnote{e-mail: mukherji@iopb.res.in}\\
\vskip .1in
{\em Institute of Physics,\\
Bhubaneswar 751~005, India.}}

\vskip .2in

{\bf Subir Mukhopadhyay\footnote{e-mail: subir@iopb.res.in}\\
\vskip .1in
{\em NISER, Institute of Physics campus,\\
Bhubaneswar 751~005, India.}}

\end{center}

\begin{center} {\bf ABSTRACT}

\end{center}
\begin{quotation}\noindent
\baselineskip 15pt
After reviewing the thermodynamics and critical phenomena associated with AdS black 
holes carrying multiple R-charges in various dimensions, we do a Bragg-Williams like
analysis of the systems around its critical points. This leads us to propose  an
effective potential governing the equilibrium properties of the boundary gauge theory.
We also study certain non-equilibrium phenomena 
associated with these gauge theories. In particular, we  compute the conductivities 
and diffusion coefficients for theories with multiple R-charges in four, three and six 
dimensions.

\end{quotation}
\vskip 2in
June 2009\\
\end{titlepage}
\vfill
\eject

\setcounter{footnote}{0}
\section{Introduction}
Recently there appear stimulating studies concerning spinning 
near-extremal D3 branes 
\cite{Son:2006em,Gubser:1998jb,Chamblin:1999tk,Cai:1998ji,Gubser:2000ec,Mas:2006dy,Maeda:2008hn,Maeda:2006by,Saremi:2006ep}. 
There are at least two reasons that motivate these activities. 
The more important one is, perhaps, the fact that this provides a simple 
example in the framework of gauge-gravity duality where different aspects 
can be studied in presence of non-zero charge as well as chemical 
potential. Indeed, the ten dimensional gravitational background 
corresponding to near extremal spinning D3 brane provides gravity dual of 
a four dimensional ${\cal{N}} = 4$ finite temparature super 
Yang-Mills theory. The number of 
independent angular momenta of the spinning D3 brane is the rank of the 
isometry group $SO(6)$ of the space tranverse to the brane, which upon 
dimensional reduction corresponds to charged black hole 
with three $U(1)$ charges. In the gauge theory, this isometry group 
represents the R-symmetry group. Three independent chemical potentials can 
therfore be introduced which couple to the three independent $U(1)$ 
charges. By exploiting the gauge-gravity correspondence, several 
recent works explored the behaviour of such gauge theories at 
strong coupling in its simplest setting. These are the cases where only one of 
the chemical potentials is non-zero.

Another reason that makes the study of rotating D3 brane  interesting is 
the fact that spinning D3 
brane has an upper critical limit of angular momentum beyond which 
thermodynamic instability sets in. Though the final configuration to 
which this instability leads is yet to be understood, it is 
conjectured that when angular momentum reaches this critical value, the 
system undergoes a phase transition. A careful inspection of the 
bahaviour of various thermodynamic quantities suggests that it is a 
second order phase transition. In particular, near this critical point, 
divergence of correlation length shows up in the behaviour of 
thermodynamic quantities such as specific heat and non-equilibrium 
quantities such as, transport coefficients. One expects that, in the 
vicinity of the second order phase transition, certain physical 
properties of the system can be extracted without the knowledge of their microscopic 
behaviours. 

With these motivations, in this paper, we study various aspects 
of the R-charged black holes near its critical point. So far these 
studies were carried out for the simplest variant of this 
model, namely, D3 branes with a single R-charge. The purpose of 
this article is to extend these stuides to its full generality. This 
leads to more intricate phase structures of the systems along with 
interesting surprises. So far the equlibrium thermodynamics is concerned, 
we carry out a Bragg-Willinas like analysis in the bulk. More specifically, by 
identifying the horizon radius as the order parameter, we write down the 
Bragg-Williams \cite{cl} potential for the R-charged black hole. 
We also extend our analysis for charged black holes in four 
and seven dimensions. They appear as the near horizon limits of 
non-extremal spinning M2 and M5 branes. Subsequently, by exploiting the 
gauge-gravity duality, we illustrate with an explicit example, how this 
leads to a proposal of effective potential for the gauge theory. As far as the 
non-equilibrium 
phenomena are concerned, we study the transport 
coefficients of the gauge thery duals using gauge-gravity correspondence. 
Besides giving a general precription to solve perturbed gravity equations
for multiple R-charged black holes in four, five and seven dimensions, we 
explicitly compute the diffusion coefficients for all these cases using 
the Green-Kubo formula.

The paper is structured as follows. In the next section, we reconsider  
various R-charged black holes, in the grand canonical enseble, with a focus on their thermodynamic 
properties \cite{Gubser:1998jb}. On one hand, this section allows us to set our notations and 
conventions and on 
the other 
hand, it  helps us to scan the complicated phase structure associated with the black 
holes with multiple R-charges in various dimensions. Next, in section 3, we consider 
the black holes near their critical temperatures. By  identifying horizon radius as 
the order parameter, we construct the  Bragg-Williams like potentials for the black 
holes. This further leads us to propose an effective potential 
for the boundary gauge theory with charge density as the order parameter. We illustarate 
this explicitly for the four dimensional gauge theory with a single non-zero   
chemical potential.
Section 4 is devoted to a study of certain non-equilibrium phenomena  
associated with these bulk geometries and its imprint on the boundary theories.  In 
particular, we compute transport coefficients using Green-Kubo formula. It turns out 
that for multiple R-charges, the calculations are quite involved. We, therofore, in the main text 
restrict ourselves to the evaluation of
the R-charge conductivity for black holes when two charges are non-zero deferring 
rest of the cases to the appendix. The appendix also 
includes an algorithm to compute the conductivities in various dimensions 
and for multiple charges. Finally, we conclude the paper in section 5 with a 
discussion on our results.
  
\section{Black Holes in STU model}

In this section we review thermodynamics of the R-charged black hole in 
five, four and seven dimensions. We start with the black holes in  
five dimensions. These are the solutions of equations 
of motion of $D =  5$, ${\cal N} = 2$ gauged supergravity obtained by 
compactification of ten 
dimensional IIB supergravity on $S^5$. These are known as STU black 
holes and were found in \cite{Behrndt:1998jd,Cvetic:1999xp,Kraus:1998hv}.
The effective action is given by
\begin{equation}
S = \frac{1}{16 \pi G} \int {\sqrt{-g}} d^5x \Big( R + \frac{2}{l^2}{\cal V} + 
\frac{1}{2} 
G_{ij} F_{\mu\nu}^i
F_{\mu\nu}^j - G_{ij} \partial_\mu X^i \partial^\mu X^j + \frac{24}{ {\sqrt{-g}}}
\epsilon^{\mu\nu\rho\sigma\lambda} \epsilon_{ijk} F_{\mu\nu}^i F^{\rho\sigma j} 
A_\lambda^k\Big),
\label{gaugeac}
\end{equation}
where $l$ represents the scale associated with the cosmological constant.
In addition to the metric, we have three scalar fields $X^i$, $i= 1,2,3$ 
which are constrained through the relation $X^1 X^2 X^3 = 1$.
${\cal V}$ is a potential involving the scalar fields given by
\begin{equation}
{\cal V} = 2 \sum_{1}^{3} \frac{1}{X^i}.
\end{equation} 
$G_{ij}$ represents metric of the moduli space parametrized by the scalars
and is given by
\begin{equation}
G_{ij} = \frac{1}{2} {\rm {diag}}\Big[ (X^1)^{-2}, (X^1)^{-2}, (X^1)^{-2}\Big].
\end{equation}
$F_{\mu\nu}^i$ are the field strengths associated with three 
abelian gauge fields $A^i$ corresponding to three $U(1)$
in the cartan of the R-symmetry group.

As shown in  \cite{Behrndt:1998jd}, this effective action (\ref{gaugeac}) 
admits asymptotically AdS black hole solutions with three $U(1)$ charges. 
These solutions can be written down as 
\begin{equation}
ds^2 = -{\cal H}^{-\frac{2}{3}} \frac{r^2}{l^2} f dt^2 + {\cal 
H}^{\frac{1}{3}}\frac{l^2}{r^2} dr^2 + {\cal H}^{\frac{1}{3}} \frac{r^2}{l^2}
(dx^2 + dy^2 + dz^2),
\label{stumet}
\end{equation}
where
\begin{equation}
f = (1 + \frac{q_1}{r^2})(1 + \frac{q_2}{r^2}) (1 + \frac{q_3}{r^2}) - \frac{r_0^4}{r^4},
\label{stumet2}
\end{equation}
and the harmonic functions are given by
\begin{equation}
{\cal H} = H_1 H_2 H_3 = (1 + \frac{q_1}{r^2})(1 + \frac{q_2}{r^2}) (1 + 
\frac{q_3}{r^2}). \label{stumet3}
\end{equation}
In the above solutions, the parameter $r_0$ is related to the mass or 
energy density of the black hole 
while $q_1, q_2, q_3$ are related to the charges of the black hole. 

\subsection{Black hole thermodynamics and instabilities in five dimensions}

In this subsection we briefly 
describe the thermodynamics and the critical phenomena associated with 
the five dimensional R-charged black hole given by 
(\ref{stumet},\ref{stumet2},\ref{stumet3}). From now, instead of using 
$r_0$ as a parameter, we 
will use the radius of the outer horizon $r_+$ to parametrize the 
solution. $r_+$ can be obtained from the 
largest root of the equation $f(r_+)=0$, where $f(r)$ is given in 
(\ref{stumet2}). Furthermore, we will scale all
the dimensionful parameters with AdS length scale $l$ e.g. we introduce 
the dimensionless parameter ${\bar r} = r_+/l$ and ${\bar T} = l T$  for 
horizon radius and Hawking temperature respectively. Similarly the 
dimensionless 
charge parameters will be denoted by ${\bar q}_i = q_i/l^2$. In terms of 
these  scaled parameters, horizon area becomes proportional to
\begin{equation}
A = \sqrt{(\bar r^2 + \bar q_1)(\bar r^2 + \bar q_2)(\bar r^2 + \bar q_3)}.
\end{equation} 
In addition, we will set Newton's constant $G=1$. $G$ and $l$ will be 
resurrected whenever required. In what follows we will consider the 
system in the grand 
canonical ensemble with temparature $T$ and chemical potentials $\mu_i$, 
$i=1,2,3$ as external parameters.

The scaled Hawking temperature turns out to be
\begin{equation}
\bar T = \dfrac{   2{\bar r}^6 + {\bar r}^4({\bar q}_1+{\bar q}_2+{\bar 
q}_3) - {\bar q}_1{\bar q}_2{\bar q}_3	}
{2\pi {\bar r}^2  A }.
\label{thermo1}
\end{equation}
Other thermodynamic quantities such as entropy, energy and pressure are expressed as
\begin{equation}
{\bar S} = \dfrac{1}{4} A ~ ,
\quad\quad
{\bar E} = \dfrac{3}{16\pi{\bar r}^2} A^2 ~ ,
\quad\quad
{\bar P} = \dfrac{1}{16\pi{\bar r}^2} A^2~,
\label{thermo2}
\end{equation}
where the above quantities are rendered dimensionless by multiplying appropriate power of $l$.
The components of the charge density and chemical potential are given by
\begin{equation}
{\bar \rho}_i = \dfrac{\sqrt{2{\bar q}_i}}{16\pi {\bar r}} A,
\quad
{\bar \mu}_i = \dfrac{\sqrt{2{\bar q}_i} }{{\bar r}({\bar r}^2 + {\bar q}_i)} A.
\label{thermo3}
\end{equation}
Since we are analysing black holes with flat horizon, the horizon has 
infinite volume. Consequently, only the themodynamic 
densities are finite. What we have written above actually represent those
thermodynamic densities.

Unless charges satisfy certain constraints, these black holes undergo a local 
instability \cite{Gubser:1998jb,Chamblin:1999tk,Cvetic:1999rb}. While at   
high temperature, black holes remain stable, once we 
reduce the temperature down to a critical value, the specific heat and 
suceptibility  diverge. In order to see this, let us compute those quantities.
The specific heat associated with the black holes has the following form
\begin{equation}
\begin{split}
&\bar C = \Big(\bar T \frac{\partial \bar S}{\partial \bar T}\Big)_{\bar \mu_1, 
\bar\mu_2, \bar \mu_3} =
(2\bar r^6+\bar r^4(\bar q_1+\bar q_2+\bar q_3)-\bar q_1\bar q_2\bar q_3)\times\\
&\dfrac{3\bar r^6-\bar r^4(\bar q_1+\bar q_2+\bar q_3)-\bar r^2(\bar q_1\bar 
q_2+\bar q_2 \bar q_3+\bar q_3 \bar q_1) + 3 
\bar q_1 \bar q_2\bar q_3}{4A ( 2\bar r^6 - \bar r^4(\bar q_1+\bar q_2+\bar 
q_3)+\bar q_1\bar q_2\bar q_3)}.
\end{split}
\end{equation}
The susceptibility, for the present case is a $3\times 3$ symmetric matrix, defined as
\begin{equation}
\bar \chi_{ij} = \Big(\frac{\partial \bar \rho_i}{\partial \bar \mu_j}\Big)_{\bar T, 
\bar \mu_k, \bar \mu_l},~{\rm with} ~~ k, l \ne j.
\end{equation}
Two of its components are given by
\begin{equation}
\begin{split}
\bar \chi_{11}
&= \dfrac{2\bar r^8 + 5\bar r^6\bar q_1-\bar r^4\bar q_1^2-\bar 
r^6\bar q_2-\bar r^4\bar q_1\bar q_2-\bar 
r^6\bar q_3-\bar r^4\bar q_1\bar q_3-3\bar r^2\bar q_1\bar q_2\bar q_3+\bar 
q_1^2\bar q_2\bar q_3}{16\pi (2\bar r^6-\bar r^4(\bar q_1+\bar q_2+\bar q_3)+\bar 
q_1\bar q_2\bar q_3)},\\ 
\bar \chi_{12}&= 
-\dfrac{\sqrt{\bar 
q_1\bar q_2}}{8\pi}\dfrac{\bar r^4 
\bar q_1+ \bar r^4\bar q_2-\bar r^4 \bar q_3-\bar 
q_1\bar q_2\bar q_3}{2\bar r^6-\bar r^4(\bar q_1+\bar q_2+\bar q_3)+\bar q_1\bar q_2\bar q_3}.
\end{split} 
\end{equation} 
The other components can be obtained by cyclically permuting the indices
$(123)$.
We now note that the specific heat and the suceptibility components diverge over the 
critical 
hypersurface
\begin{equation}
2\bar r^6 - \bar r^4(\bar q_1+\bar q_2+\bar q_3)+\bar q_1\bar q_2\bar q_3 = 0.
\label{hype}
\end{equation}

The thermodynamic quantities for a subclass of black holes where one or more charges are 
turned off can be easily obtained from the above by appropriate  
substitutions.
Apart from some isolated cases, which we will discuss in the sequel, all these black holes show 
instabilities. The only difference is that hypersurfaces at which divergences 
occur are different from (\ref{hype}) and depend on the nature of the black hole. This is 
listed below for black holes with single, double or three charges :
\begin{eqnarray}
&&2 \bar r^2 - \bar q_1 =0, \qquad {\rm for} ~\bar q_1 \ne 0, \bar q_2 = \bar q_3= 
0,\nonumber\\
&& 2 \bar r^2 - \bar q_1 - \bar q_2 = 0, \qquad {\rm for} ~\bar q_1, 
\bar q_2 \ne 0, \bar q_3 = 0, \nonumber\\
&& 2\bar r^4 - \bar q_1 \bar r^2 - \bar q_1 \bar q_2 = 0,  \qquad {\rm for} ~\bar q_1 
\ne 
0, \bar q_2 = \bar q_3 \ne 0.
\label{hypetwo}
\end{eqnarray}
We also note that thermodynamic quantities are everywhere well behaved in  
the cases when charges satisfy certain restrictions. These include $\bar 
q_1 = \bar q_2$, $\bar q_3 = 0$ and 
$\bar q_1 = \bar q_2 = \bar q_3$. The second one is the standard Reissner-Nordstrom 
black hole in AdS space. Theromodynamic properties of these black holes were discussed 
in detail in \cite{Chamblin:1999tk}. The critical surfaces given in 
(\ref{hypetwo}) can be expressed 
as relations between temperature and chemical potentials. For example, 
for one and two charge black holes, they are given respectively by
\begin{eqnarray}
&& {\bT} = \frac{\bmu_1}{\pi},\nonumber\\
&&\bar T = \frac{\{\bar \mu_1^2 + \bar \mu_2^2 + (\bar \mu_1 - \bar\mu_2)^{\frac{4}{3}}
(\bar \mu_1 + \bar\mu_2)^{\frac{2}{3}} + (\bar \mu_1 - \bar\mu_2)^{\frac{2}{3}}\bar \mu_1 
+
\bar\mu_2)^{\frac{4}{3}}\}^{\frac{3}{2}}}
{\pi {\sqrt{ 4 \bar \mu_1^2 \bar \mu_2^2 + 3 \{\bar \mu_1^2 + \bar \mu_2^2 + (\bar \mu_1 
-
\bar\mu_2)^{\frac{4}{3}}
(\bar \mu_1 + \bar\mu_2)^{\frac{2}{3}} +  (\bar \mu_1 - \bar\mu_2)^{\frac{2}{3}}\bar 
\mu_1 +
\bar\mu_2)^{\frac{4}{3}}\}^{2}}}}. \nonumber\\
\nonumber\\
\label{csurf}
\end{eqnarray}
For a generic three charged black hole, we have not been able to find the critical surface 
in terms of temperature and chemical potentials in a compact form. For a given set of 
$\bar \mu_i$, (\ref{csurf}) gives us the critical temperature at which black hole 
instability arises. As we do not yet know the stable configuration below 
this temperature, in this paper we consider the black holes slightly {\it 
above} the critical temperature. 

As one approaches  the critical surface the correlation length diverges. 
This shows up in the divergences of thermodynamic quantities. Near the 
critical temperature, these black holes have universal features and their 
behaviour do not depend on details of the theory. These features are 
encoded in the critical exponents associated with behaviour of various 
thermodynamic quantities as we approach the critical surface.

We begin with the following four static critical exponents.
Approaching the critical surface with fixed $\mu_i$, one finds three critical exponents
$\alpha, \beta, \gamma$ associated with the bahaviour of specific heat, charge densities and susceptibilities respectively.
\begin{equation}
\bar C \sim ( \bar T - \bar T_c)^{-\alpha}, \bar \rho - \bar \rho_c \sim ( \bar T - \bar 
T_c)^{\beta}, \bar \chi \sim ( \bar T - \bar T_c)^{-\gamma}.
\end{equation}
 Since there are three $U(1)$'s, there are three charge densities and so the number of respective exponents $\beta_i$ is also three. However, it turns out that they are all equal and one gets only a single $\beta$. Similarly, the susceptibility matrix $\chi_{ij}$ being symmetric and real, can be diagonalized as $\chi_{ij} = \chi_i\delta_{ij}$ and generically one would expect three $\gamma$'s one for each of the components. But since all the three components are equal we get one $\gamma$ only. 
On the other hand, exponent $\delta$ is obtained by approaching the critical line 
with a trajectory on which the temperature is constant. This is defined as
\begin{equation}
\bar \rho - \bar \rho_c \sim - (\bar \mu - \bar \mu_c)^{\frac{1}{\delta}}.
\end{equation}
We find for all the black holes which show instability have the same set of
critical exponents
\begin{equation}
(\alpha, \beta, \gamma, \delta) = (\frac{1}{2}, \frac{1}{2}, \frac{1}{2}, 2).
\label{expofive}
\end{equation}
These critical exponents are not independent but are related through scaling relations. These critical exponents satisfy
\begin{equation}
\alpha + 2 \beta + \gamma = 2, ~~\gamma = \beta (\delta -1).
\end{equation}

There are two other critical exponents $\nu$ and $\eta$ associated with the behaviour of correlation length and correlation function near critical surface. If $G(\vec r)$ is the correlator its behaviour near the critical surface is
\begin{equation}
\begin{split}
G(\vec r) &\sim e^{-r/\xi}\quad \text{at}\quad T \neq T_c, 
\quad \xi \sim \left(\dfrac{T-T_c}{T_c}\right)^{-\nu}\\
G(\vec r) &\sim r^{-d+2-\eta},\quad \text{at}\quad T = T_c.
\end{split}
\end{equation}
Assuming additional scaling relations $\gamma = \nu(2-\eta)$ and $2-\alpha = \nu d$, to be valid we can determine $\nu$ and $\eta$ to be equal to $\dfrac{1}{2}$ and $1$ respectively. However, it is not clear a priory whether the scaling relations are valid. One can compute $\nu$ and $\eta$ from correlation of scalar modes in the gravity theory. However, if we assume the scaling relations are valid then this will belong to class of B-model in classification given in \cite{Hohenberg:1977ym}.

As we will see in the next section, it is possible to construct a Bragg-William energy
function which reproduces all these critical exponents near the second order phase 
transition surfaces. This function, in turn, will allow us to propose an effective 
potential for the boundary gauge theory via AdS/CFT 
correspondence.

\subsection{Black Hole thermodynamics and instabilities in four and seven 
dimensions}

In this subsection, we briefly discuss the thermodynamics and  
instabilities of R-charged 
black holes in four and seven dimensions. As mentioned earlier, these 
configurations are 
obtained from spinning $M2$ and $M5$ branes of eleven dimension supergravity.
First, let us focus on the black holes in four dimensions. They can carry atmost four 
independent charges arising form Cartans of $SO(8)$. The black hole solutions are 
discussed in \cite{Cvetic:1999xp}.
For our purpose, we require the thermodynamic quantities only. Our notation remains
the same as in the last subsection except now there will be four charges $q_i$, $i=1,2,3,4$.
For convenience, once again we introduce the scaled variables ${\bar r}$ and ${\bar q}_i$
and the horizon area is proportional to
\begin{equation}
A=\sqrt{(\bar r + \bar q_1^2)(\bar r + \bar q_2^2)(\bar r + \bar q_3^2)(\bar r + \bar q_4^2)}.
\end{equation}
The temperature is given by
\begin{equation}
\bar T = \left( - \frac{1}{\bar r} + \sum_{j = 1, 4} \frac {1}{\bar r + \bar q_j^2}\right) A.
\end{equation}
The entropy, energy and pressure are given by
\begin{equation}
{\bar S} = \dfrac{1}{4} A~,
\quad\quad
\bar E = \frac{1}{8 \pi \bar r} A^2~,
\quad\quad
\bar P = \frac{1}{36 \pi \bar r} A^2 ~ .
\end{equation}.
The components of charge and chemical potentials are 
\begin{equation}
\bar \rho_i = \frac{1}{32 \pi}\sqrt{ \frac{2 \bar q_i^2}{\bar r} }A~,
\quad\quad
\bar \mu_i = {\sqrt{ \frac{2 \bar q_i^2 }{\bar r (\bar r + \bar q_i^2)^2}}} A~.
\label{thermofour}
\end{equation}

The specific heat is given by
\begin{equation}
\bar C = \Big(\bar T \frac{\partial \bar S}{\partial \bar T}\Big)_{\bar\mu_1, 
\bar\mu_2, \bar\mu_3, \bar\mu_4}
= \frac{c_1 c_2} { 4 \Delta A},
\end{equation}
where $c_1, c_2, \Delta$ are given by
\begin{equation}
\begin{split}
&c_1 = 2 \bar r^4 - (\sum_{i=1,4} \bar q_i^2) \bar r^3 +
( \bar q_1^2 \bar q_2^2 \bar q_3^2 + \bar q_2^2 \bar q_3^2  \bar q_4^2
+  \bar q_3^2  \bar q_4^2 \bar q_1^2 + \bar q_4^2 \bar q_1^2 \bar q_2^2) \bar r
- 2 \prod_{i=1,4} \bar q_i^2, \\
&c_2 = 3 \bar r^4 + 2 (\sum_{i=1,4} \bar q_i^2) \bar r^3 + ( \bar q_1^2 \bar q_2^2 +
 \bar q_1^2 \bar q_3^2 +  \bar q_2^2 \bar q_3^2 + \bar q_1^2 \bar q_4^2 +  \bar q_2^2 
\bar q_4^2  +  \bar q_3^2 \bar q_4^2 )\bar r^2 - \prod_{i=1,4} \bar q_i^2,\\
&\Delta = 3 \bar r^4 - 2 (\sum_{i=1,4} \bar q_i^2) \bar r^3 + (\bar q_1^2 \bar q_2^2 
+ \bar q_2^2 \bar q_3^2 + \bar q_3^2  \bar q_4^2  +  \bar q_1^2 \bar q_3^2 + \bar q_1^2 
\bar q_4^2 + \bar q_2^2  \bar q_4^2) \bar r^2 - \prod_{i=1,4} \bar q_i^2.
\label{m2crit}
\end{split}
\end{equation}
The suceptibility in this case is a four dimentional 
metrix defined as 
\begin{equation}
\bar \chi_{ij} = \Big(\frac{\partial \bar \rho_i}{\partial \bar \mu_j}\Big)_{\bar T,
\bar \mu_k, \bar \mu_l, \bar \mu_m},~{\rm with} ~~ k, l, n \ne j.
\end{equation}
Two of the components are 
\begin{eqnarray}
\bar \chi_{11} &&= \Big(\frac{\partial \bar \rho_1}{\partial \mu_1}\Big)_{\bar T, 
\bar\mu_2, \bar\mu_3, \bar\mu_4} = \frac{\bar r \delta_1}{32 \pi 
\Delta},\nonumber\\
\bar \chi_{12} &&= \Big(\frac{\partial \bar \rho_1}{\partial \mu_2}\Big)_{\bar T,
\bar\mu_1, \bar\mu_3, \bar\mu_4} = -\frac{ \bar q_1 \bar q_2 \delta_2}{ 32 \pi 
\Delta},
\end{eqnarray}
where
\begin{equation}
\delta_1 = 3 \bar r^4 + 3 ( 2 \bar q_1^2 - \bar q_2^2  - \bar q_3^2 - \bar q_4^2) 
\bar r^3 + ( \bar q_2^2 \bar q_3^2 +  \bar q_3^2 \bar q_4^2  + \bar q_2^2 \bar q_4^2  
- 3( \bar q_1^2 \bar q_2^2 - \bar q_1^2 \bar q_3^2 - \bar q_1^2 \bar q_4^2) )
+ 3\prod_{j = 1,4} \bar q_j^2,\nonumber
\end{equation}
and
\begin{equation}
\delta_2 = \bar r^4 + 2 (\bar q_1^2 + \bar q_2^2  - \bar q_3^2 - \bar q_4^2)
\bar r^3 + ( 3 \bar q_3^2 \bar q_4^2 
- \bar q_1^2 \bar q_2^2 - \bar q_1^2 \bar q_3^2 - \bar q_2^2 \bar q_3^2 
- \bar q_1^2 \bar q_4^2 - \bar q_2^2 \bar q_4^2) \bar r^2 + 
\prod_{j = 1,4} \bar q_j^2.\nonumber
\end{equation}
Other components of suceptibility can be obtained by cyclically permuting the indices
$(1234)$. We note that specific heat and suceptibility diverge at $\Delta 
= 0$. 

Apart from some isolated cases (discussed below), 
all black holes suffer from instabilities. When some of the charges are 
either equal or zero, the critical surfaces change as can be seen directly by 
sustituating charges in the general expressions above. We list down various such critical 
surfaces below when some of the charges are identically zero: 
\begin{eqnarray}
&&3 \bar r - 2 \bar q_1^2 =0, \qquad {\rm for} ~\bar q_1 \ne 0, \bar q_2 = \bar q_3 = 
\bar q_4 = 0, \nonumber\\
&& 3 \bar r^2 - 2 \bar r (\bar q_1^2  + \bar q_2^2) + \bar q_1^2 \bar q_2^2 =0, \qquad  
{\rm for} ~\bar q_1, \bar q_2 \ne 0, \bar q_3 = \bar q_4 = 0.
\end{eqnarray}
Furthermore, unlike the black holes in five dimensions, these black holes show 
instability even when two of their charges are equal. In particuler, $\bar q_1 =\bar q_2 
\ne 0$ and with $ \bar q_3 = \bar q_4 = 0$, the singularities appear at
\begin{equation}
3 \bar r - 2 \bar q_1^2 =0.
\label{singone}
\end{equation}
For black holes with three charges, the critical surfaces are
\begin{eqnarray}
&&3 \bar r^2 - 2 \bar r ( \bar q_1^2 + \bar q_2^2 + \bar q_3^2) + \bar q_1^2 \bar q_2^2 
+ \bar q_1^2 \bar q_3^2 + \bar q_2^2 \bar q_3^2 = 0 \qquad {\rm for}~ \bar q_1, \bar 
q_2, \bar q_3 \ne 0, \bar q_4 = 0,\nonumber\\
&& 3 \bar r - \bar q_1^2 - 2 \bar q_3^2 =0\qquad {\rm for} ~ \bar q_1 = \bar q_2 \ne 0, 
\bar q_3 \ne 0,  \bar q_4 = 0.
\end{eqnarray}
We further note that, for three non-zero charges, black holes are 
thermodynamically 
stable only when $\bar q_1 = \bar q_2 = \bar q_3$. Though for  generic values of four 
charges, instabilty appears at $\Delta =0$, when some of the charges are equal 
location of thermodynamic singularities change. This is given in the following list.
\begin{eqnarray}
&& 3 \bar r^3 - \bar r^2 ( 2 \bar q_1^2 + 2 \bar q_2^2 + \bar q_3^2 ) + \bar r  \bar 
q_1^2 \bar q_2^2 + \bar q_1^2 \bar q_2^2 \bar q_3^2 =0 \qquad {\rm for} ~\bar q_1 \ne 
\bar q_2, \bar q_3 = \bar q_4, \nonumber\\
&& 3 \bar r^3  - \bar r (\bar q_1^2 + \bar q_3^2) - q_1^2 q_3^2 = 0
\qquad {\rm for} ~\bar q_1 = \bar q_2,  \bar q_3 = \bar q_4, \nonumber\\
&& 3 \bar r^2 - 2 \bar r q_4^2 - q_1^2 q_4^2 = 0 \qquad {\rm for} ~\bar q_1 = \bar q_2 =  
\bar q_3, \bar q_4 \ne 0.
\end{eqnarray}
However, when all the charges are equal, thermodynamic instabilities are {\it absent}.

Approaching the critical surfaces in different ways, it is possible to find all the
critical exponents as in the previous subsection. In all the cases it turns out that the
exponents are same as in (\ref{expofive}).

In the rest of this subsection, we consider the thermodynamic properties of black holes in 
seven dimensions. Since those black holes correspond to $S^4$ reduction of spinning $M5$ 
branes in eleven dimensional supergravity, the Cartan consists of $U(1)\times U(1)$. 
Therefore, a general R-charged black hole can 
carry only two independent charges. The general form of the solutions are given in 
\cite{Cvetic:1999xp}. We only list down the thermodynamic variables.
The horizon area is proportional to
\begin{equation}
A = {\sqrt{(\bar r^4 + \bar q_1^2) ( \bar r^4 + \bar q_2^2)}},
\end{equation}
The temperature is
\begin{equation}
\bar T = \frac { 3 \bar r^8 + \bar r^4 (\bar q_1^2 + \bar q_2^2) - \bar q_1^2 \bar 
q_2^2}
{2 \pi \bar r^3 A}~.
\end{equation}
Entropy, energy and pressure are
\begin{equation}
\bar S = \dfrac{A}{4}~,
\quad\quad
\bar E = \frac{5 A^2}{16 \pi}~,
\quad\quad
\bar P = \frac{A}{16 \pi}.
\end{equation}
The components of the charge and chemical potential are
\begin{equation}
\bar \rho_i = \frac{ {\sqrt{2}\bar q_i}}{8 \pi\bar r}A~,
\quad\quad
\bar \mu_i = \frac{\bar q_i}{\bar r ( \bar r^4 + \bar q_i^2)} A.
\label{thermoseven}
\end{equation}
The specific heat is
\begin{equation}
\bar C = \frac{\bar r ( 5 \bar r^8 - \bar r^4 \bar q_1^2 - \bar r^4 \bar q_2^2 - 3 \bar 
q_1^2 \bar q_2^2)( 3 \bar r^8 + \bar r^4 \bar q_1^2 + \bar r^4 \bar q_2^2 - \bar
q_1^2 \bar q_2^2)}{3 \bar r^8 - \bar r^4 \bar q_1^2 - \bar r^4 \bar q_2^2 - q_1^2 \bar 
q_2^2}.
\end{equation}
The suceptibility is a two dimensional symmetric matrix with entries
\begin{eqnarray}
&&\bar \chi_{11} = \frac{\bar r( 3 \bar r^{12} + 13 \bar r^8 \bar q_1^2 - \bar r^8 \bar 
q_2^2 - 4 \bar r^4 \bar q_1^4 + \bar r^4 \bar q_1^2 \bar q_2^2 - 4 \bar q_1^4 \bar 
q_2^4)}{ 8\pi ( 3 \bar r^8 - \bar r^4 \bar q_1^2 - \bar r^4 \bar q_2^2 - \bar q_1^2 \bar
q_2^2)},\nonumber\\
&&\bar \chi_{12} = -\frac{\bar q_1 \bar q_2 \bar r( 5 \bar q_1 \bar q_2 + 
3 \bar r^4 \bar q_1^2 + 5 \bar r^4 q_2^2 - 5 \bar r^8)}
{3 \bar r^8 - \bar r^4 \bar q_1^2 - \bar r^4 \bar q_2^2 - \bar q_1^2 \bar q_2^2}.
\end{eqnarray}
$\bar \chi_{22}$ can be obtained by interchanging the indices of the expression of $\bar 
\chi_{11}$. Thermodynamic instabilities appear for all the black holes 
unless $\bar q_1 = \bar q_2$. The critical surfaces are listed below.
\begin{eqnarray}
&&3 \bar r^8 - \bar r^4 \bar q_1^2 - \bar r^4 \bar q_2^2 - q_1^2 \bar q_2^2 = 0 ~{\rm 
for} ~\bar q_1 \ne \bar q_2 \ne 0,\nonumber\\
&& 3 \bar r^4 - \bar q_1^2 = 0, ~{\rm for} ~\bar q_1 \ne 0, \bar q_2 = 0.
\end{eqnarray}
Critical exponents turn out to be same as (\ref{expofive}). 


\section{Mean Field Analysis}

In the previous section, we studied the thermodynamic properties of R-charged black 
holes in five, four and seven dimensions.
We have seen that, for generic values of charges, these black holes undergo 
continuous phase transition characterized by certain values of critical 
exponents. In this section, our aim is 
to provide a mean field description of this black hole phase transition. 
More specifically, 
within the grand canonical ensemble, by identifying the horizon radius as 
an order parameter, 
we provide Bragg-Williams like effective potentials \footnote{Similar 
constructions were also found useful in oreder to study various black holes and 
gauge theories around the Hawking-Page points. See for example \cite{Dey:2006ds, 
Dey:2007vt, Dey:2008bw}.} that are expected to decribe instabilities associated with these 
black holes\footnote{Though in this paper, we will interchangeably use the words 
Bragg-Williams 
potential and mean-field potential, it should be recognized that BW potential and Landau mean-field 
potential are not the same. Close to $T_c$, the later appears, in some cases, as 
an 
expansion of 
the former in terms of order paramer. We will comment on such a phenomenon in section 3.3.}. 
On shell, this potential reduces to the 
grand canonical 
potential. Furthermore, we will see, near the critical surfaces, the 
effective potentials 
leads to the right set of critical exponents. 

This section has three subsections. In the first subsection, 
we analyze black holes in five dimensions. In particular, we provide the effective 
potentials for black holes with one or two charges. For the case of three non-zero 
charges, owing to the complexity of the thermodynamic variables, we have not been able to 
construct mean field potential in a simple closed form. In the next subsection, we carry 
out a similar exercise for black holes in four and seven dimensions. Finally, in the
last part of this section, by trading horizon radius with charge density of the 
boundary gauge theory, we propose an effective potential which is expected to decribe the 
gauge theory in the presence of non-zero chemical 
potential. We illustrate this explicitly for simplest R-charged black 
hole, namely, for a singly charged black hole in five dimensions.

\subsection{Black holes in five dimensions}

We start with the simplest of R-charged black holes - one with a single 
non-zero charge 
in five dimensions. Thermodynamic quantities can be found out simply 
setting $\bar q_2 = 
\bar q_3 
= 0$ in (\ref{thermo1}, \ref{thermo2}, \ref{thermo3}). We suggest that it 
is described, by an effective potential \footnote{Constructional 
procedure is similar to the one described in \cite{cl}.}
\begin{equation}
\bF (\bT, \bmu_1, \br) = \frac{1}{8 \pi}\Big[\frac{\br^4 ( 3 \br^2 - \bmu_1^2)}{2 \br^2 - 
\bmu_1^2} -
\frac{2 {\sqrt 2} \pi \br^4 \bT}{\sqrt{2 \br^2 - \bmu_1^2}}\Big].
\label{poten}
\end{equation}
Note that, as in grand canonical ensemble, $\bT, \bmu_1$ are to be 
treated as external 
parameters. $\bF$ also depends upon $\br$, the dimensionless horizon radius. $\br$ will 
be treated here as the order parameter. As we will discuss below, our choice leads to 
the correct set of critical exponents near the critical line. We also point out that 
in $\bF$, $\br$ can take any value; the relation between $\br$ and 
the mass and the 
charge of the black hole appears as a consequence of on-shell condition.

The equilibrium condition of the system is represented by
\begin{equation}
\frac{\partial \bF}{\partial \br} = 0.
\end{equation}
This leads to
\begin{equation}
\bT = \frac{ (4\br^2 -\bmu_1^2)}{
2 {\sqrt{2}} \pi {\sqrt{(2 \br^2 - \bmu_1^2)}} } = \frac{\bq_1 + 2 \br^2}{2\pi 
\sqrt{\br^2 + \bq_1}}.
\label{tempback}
\end{equation}
This is the temperature of a singly charged black hole as 
can be seen from (\ref{thermo1}) after substitutions $\bar q_2 = \bar q_3 
= 0$. 
Now using (\ref{tempback}) in (\ref{poten}), we get 
\begin{equation}
\bF = \frac{\br^6}{8\pi(\bmu_1^2 - 2 \br^2)} = -\frac{\br^2 (\bq_1 + \br^2)}{16 \pi} = - 
\bar P.
\end{equation}
So on-shell, $\bF$ reduces to the negative of the pressure $\bar P$ which is same as the grand 
canonical potential. The extrema of (\ref{poten}) appear at
\begin{eqnarray}
\br &&= 0, \nonumber\\
&&= {\sqrt{\frac{2}{3}}} \bmu_1,\nonumber\\
&&= \frac{1}{2}{\sqrt{ {\bmu_1^2} + 2 \pi^2 \bT^2 - 2 {\sqrt{\pi^4 \bT^4
- \pi^2\bmu_1^2 \bT^2}}}},\nonumber\\
&&= \frac{1}{2}{\sqrt{ {\bmu_1^2} + 2 \pi^2 \bT^2 + 2 {\sqrt{\pi^4 \bT^4
- \pi^2\bmu_1^2 \bT^2}}}}.
\label{root}
\end{eqnarray}
We note that within the range of our interest, namely for $\bar q_1 \le 2 \bar r^2$ 
(which leads to $\bar T \ge \bar T_c$) ,
fourth root of (\ref{root}) corresponds to the only minimum. This equilibrium value of
$\br$ increases with the temperature.

\begin{figure}[t]
\begin{center}
\begin{psfrags}
\psfrag{a}[][]{$\br$}
\psfrag{b}[][]{$\bF$}
\epsfig{file=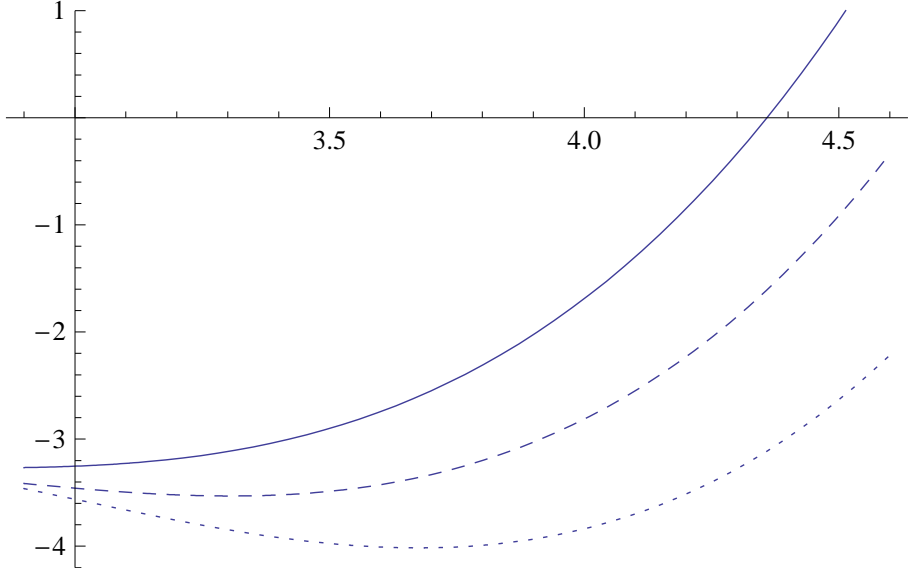, width= 12cm,angle=0}
\end{psfrags}
\vspace{ .1 in }
\caption{$\bF$ as a function of order parameter $\br$ for fixed $\bar\mu_1$. The
solid line is for the $\bar T = \bar
T_c$. The other two lines are for $\bar T > \bar T_c$. As $\bT$ increases, minimum of $\bF$
appears for larger values of $\br$. From $\bT = \bar T_c$, the order parameter changes
continuously with $\bT$.}
\label{lgone}
\end{center}
\end{figure}

Behaviour of $\bF$ as a 
function of the order parameter for fixed $\bmu_1$ and for different temperatures is shown 
in the figure (\ref{lgone}). Clearly, the order parametr $\br$ changes continuously around the 
critical temperature. This is typical of a second order phase transition.

The mean field potential $\bF$ reporduces correct set of critical exponents discussed 
in sec. 2.1. Approaching the critical line 
along constant $\bar \mu_1 = \bar \mu_{1c}$, we see from the last root of (\ref{root})
\begin{equation}
\br - \br_c \sim (\bar T - \bar T_c)^{\frac{1}{2}}, ~{\rm with} ~\br_c = \frac{\sqrt 3}{2} \bar 
\mu_{1c}.
\end{equation}
This leads to the critical exponent $\beta = \frac{1}{2}$. Similarly, the susceptibility, that 
follows from (\ref{root}), behaves as 
\begin{equation}
\bar \chi = \frac{\partial \br}{\partial \bmu_1}|_{\bar T} \sim (\bT - 
\bT_c)^{-\frac{1}{2}},
\end{equation}
near the critical temperature.
Therefore $\gamma = \frac{1}{2}$. 
The specific heat follows from (\ref{poten}). 
\begin{equation}
\bar C_{\bmu_1} = -\bT \frac{\partial^2 \bF}{\partial T^2}|_{\bar\mu_1} 
\sim (\bT - 
\bT_c)^{-\frac{1}{2}},
\end{equation}
which gives $\alpha = \frac{1}{2}$. Finally, approaching the critical line 
with 
$\bT = \bT_c$ and using (\ref{thermo3}), we get from the last root of 
(\ref{root})
\begin{equation}
\br -{\br}_c \sim -~(\bmu_{1c} - \bmu_1)^{\frac{1}{2}}, ~{\rm with} ~\br_c = 
\frac{\pi{\sqrt 3}}{2} 
\bar
T_c.
\end{equation}
where ${\br}_c$ is the critical value of $\br$ at $\bmu_1 = {\bmu}_{1c}$.
Hence, we have $\delta = 2$. We therefore conclude that the effective potential of 
(\ref{poten}) reproduces all the critical exponents given in (\ref{expofive}).

Next, we consider black holes with two non-zero charges. Thermodynamic 
variables can be 
found from (\ref{thermo1},\ref{thermo2},\ref{thermo3}) by substituting 
$\bar q_3 = 0$. The mean 
field potential 
can be expressed in the following manner:
\begin{equation}
\bF (\bar T, \bar \mu_1, \bar \mu_2, \bar r) = 
\frac{(\bar\mu_1^2 \bar\mu_2^2 - 4 \bar r^4)}{8 \pi}\Big[ \frac{
(12 \bar r^4 -4 \bar\mu_1^2 \bar r^2 - 4 \bar\mu_2^2 \bar r^2 + \bar\mu_1^2 \bar\mu_2^2)}
{8 (\bar \mu_1^2 - 2 \br^2)(\bar \mu_2^2 - 2 \br^2)}
- \frac{\bar T \bar r \pi}
{\sqrt{(\bar \mu_1^2 - 2 \br^2)(\bar \mu_2^2 - 2 \br^2)}}\Big].
\label{twochargepot}
\end{equation} 
The potential is a function of the order parameter $\br$ and also depends on
the external parameters $\bar T, \bar \mu_1$ and  $\bar \mu_2$. Note that in the 
limit $\bar\mu_2 =0$, $\bF$ reduces to that of (\ref{poten}).

The equilibrium configuration is found by minimizing the potential with respect to
$\br$. This leads to 
\begin{equation}
\bar T = \frac{\br (4 \br^2 - \bar\mu_1^2 - \bar \mu_2^2)}{ 2 \pi {\sqrt
{(  \bar\mu_1^2 - 2 \br^2) (  \bar\mu_2^2 - 2 \br^2)}}} = 
 \frac{\bar r}{2\pi} \frac{\bar q_1 + \bar q_2 + 2 \bar r^2}
{\sqrt{(\bar r^2 + \bar q_1)(\bar r^2 + \bar q_2)}}.
\label{tempmean}
\end{equation}
This is the equilibrium temperature of the doubly charged black hole 
(\ref{thermo1}). 
Substituting (\ref{tempmean}) in (\ref{twochargepot}), we get
\begin{equation}
\bF = - \frac{(\bar\mu_1^2 \bar\mu_2^2 - 4 \br^4)^2}{64 \pi ( \bar\mu_1^2 - 2 \br^2)(  
\bar\mu_2^2 - 2 \br^2)} = 
-\frac{1}{16 \pi} (\bar r^2 + \bar q_1)(\bar r^2 + \bar q_2) = -\bar P.
\end{equation}

\begin{figure}[t]
\begin{center}
\begin{psfrags}
\psfrag{a}[][]{$\br$}
\psfrag{b}[][]{$\bF$}
\epsfig{file=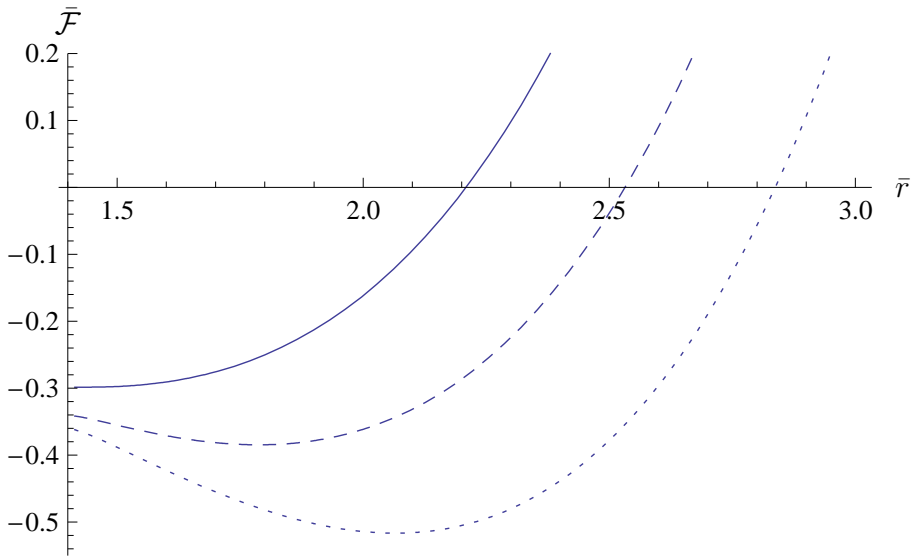, width= 12cm,angle=0}
\end{psfrags} 
\vspace{ .1 in }
\caption{$\bF$ as a function of order parameter $\br$ for fixed $\bar\mu_1$ and $\bar\mu_2$. The 
solid line is for the $\bar T = \bar 
T_c$. The other two lines are for $\bar T > \bar T_c$. As $\bT$ increases, minimum of $\bF$ 
appears for larger values of $\br$. From $\bT = \bar T_c$, the order parameter changes 
continuously with $\bT$} 
\label{lgonetwo}
\end{center}
\end{figure}

The values of $\br$ at which $\bF$ has extrema can be evaluated analytically. However, 
expressions are very large and unilluminating. We do not display them here.
It turns out that in the region of our interest namely for $2 \br^2 - \bar q_1 - 
\bar q_2 \ge 0$, there is only a single minimum of $\bF$. Typical behaviour of $\bF$ as 
a function of $\br$ for fixed $\bar\mu_1$ and $\bar\mu_2$ is shown in the figure 
(\ref{lgonetwo}). We see that
the order parameter $\br$, at which the  minimum appears, continuously increases as we increase 
the temperature. This shows that the phase transition is of second order. 

Various critical exponents can be evaluated as in the single charged case. Approaching
the critical surface along constant $\bar\mu_1$ and $\bar\mu_2$, and expanding the 
minimum value of $\br$ close to the critical temperature, we get 
\begin{equation}
\br - \br_c \sim (\bar T - \bar T_c)^{\frac{1}{2}}.
\end{equation}
Similarly, on constant $\bar T, \bar\mu_2$ and $\bar T, \bar\mu_1$ surfaces, we get
\begin{equation}
\bar \chi_1 = \frac{\partial \br}{\partial \bar\mu_1} \sim (\bar T -\bar T_c)^{-\frac{1}{2}},
\qquad \bar \chi_2 = \frac{\partial \br}{\partial \bar\mu_2} \sim (\bar T -\bar 
T_c)^{-\frac{1}{2}},
\end{equation}
respectively. The specific heat follows from taking second derivative of $\bF$ with respect to
$\bar T$ for fixed $\bar \mu_1$ and $\bar \mu_2$. This leads to
\begin{equation}
C_{\bar\mu_1, \bar\mu_2} = -\bT \frac{\partial^2 \bF}{\partial \bar T^2} \sim (\bar T - \bar 
T_c)^{-\frac{1}{2}}.
\end{equation}
Finally, on constant  $\bar T, \bar\mu_2$ and $\bar T, \bar\mu_1$ surfaces, expanding around the 
minimum  of $\bF$ for $\bar \mu_1$ and $\bar\mu_2$  close to their critical values, we get
\begin{equation}
\br - \br_c \sim - ~(\bar \mu_1 - \bar \mu_{1c})^{\frac{1}{2}}, \qquad 
\br - \br_c \sim - ~(\bar \mu_2 - \bar \mu_{2c})^{\frac{1}{2}},
\end{equation}
respectively. We now note that these lead to the same critical exponents that we got from our 
analysis in the previous section, see (\ref{expofive}).

\subsection{Black holes in four and seven dimensions}

Here we reconsider the black holes in four and seven dimensions. As in previous 
subsection, our aim will be to understand their behaviours near the critical surfaces in 
terms of effective potentials. Though for all non-zero charges, it is in 
principle 
possible to construct the function $\bF$, it becomes computationally 
difficult to express 
$\bF$ in terms of temperature and all non-zero chemical potentials in a 
closed form. We 
will therefore 
consider, in this subsection, black holes with a single charge.

Thermodynamic variables for four dimensional holes with single charge can 
be found out by 
substituing $\bar q_2 = \bar q_3 = \bar q_4 = 0$ in (\ref{thermofour}). The effective 
potential can be written down as
\begin{equation}
\bF (\bar T, \bar \mu_1) = \frac{1}{16 \pi} \bigg[ \frac{\bar r^3 ( 4 \bar r^2 -\bar 
\mu_1^2)}{ 2 \bar r^2 - \bar\mu_1^2} - 4 {\sqrt 2} \pi \bar T \bar r^3 {\sqrt
{ \frac{1}{ 2 \bar r^2 - \bar \mu_1^2}}}\bigg].
\label{lgfour}
\end{equation}
The saddle point of  $\bF$ gives the on-shell temperature of the black hole
\begin{equation}
\bar T = \frac{6 \bar r^2 - \bar\mu_1^2}{4 {\sqrt 2} \pi {\sqrt {2\bar r^2 - \bar \mu_1^2}}} 
= \frac{(3 \bar r + 2 \bar q_1^2) {\sqrt {\bar r}}}{ 4 \pi {\sqrt {\bar r + \bar q_1^2}}}.
\end{equation}
This is same as what we would have gotten from (\ref{thermofour}) after 
setting three charges to zero. 

The critical line (\ref{singone}), while expressed in terms of $\bar T$ and $\bar \mu_1$ 
reads as $\bar T_c = {\sqrt 3} \bar\mu_1/2\pi$. We will be interested in the rigion 
$\bar T \ge \bar T_c$ or equvalently in the region $\bar q_1^2/r \le 3/2$. There is only 
one minimum of $\bF$ in this region and is given by
\begin{equation}
\bar r = {\sqrt{ \frac{\bar\mu_1^2}{6} + \frac{8 \pi^2 \bar T^2}{9} +
\frac{4}{9} {\sqrt{ 4 \pi^4 \bar T^4 - 3 \pi^2 \bar \mu_1^2 \bar T^2}}}}.
\end{equation}
Studying $\bF$ around this point, as in the last subsection, we recover the critical 
exponents (\ref{expofive}).

For seven dimensional black holes, one can similarly construct the effective potential 
$\bF$. Following is the expression of $\bF$ for single charge case.
\begin{equation}
\bF (\bar r, \bar \mu_1) = \frac{\bar r^6}{8 \pi}\bigg[ \frac{5 \bar r^2 - 2 \mu_1^2}{2 
\bar r^2 - \mu_1^2} - 2 {\sqrt 2}\pi \bar T {\sqrt{\frac{1}{2 \bar r^2 - \mu_1^2}}}\bigg].
\label{lgseven}
\end{equation}
As before at the saddle point this function reduces to grand canonical potential with 
temperature given by the one in (\ref{thermoseven}) with $\bar q_2 = 0$. The critical 
surface appears at $\bar T = {\sqrt{\frac{3}{2}}}\frac {\bar \mu_1}{\pi}$. For a given 
$\bar\mu_1$, this equation defines $\bar T_c$. In the range of our interests, namely for 
$\bar T \ge \bar T_c$, there is only one minimum of $\bF$. This is given by
\begin{equation}
\bar r = {\sqrt{  \frac{\mu_1^2}{3} + \frac{2 \pi^2 \bar T^2}{9} + \frac{1}{9}
{\sqrt{ 4 \pi^4 \bar T^4 - 6 \pi^2\bar \mu_1^2 \bar T^2}} }}.
\end{equation}
Studying the behaviour of $\bF$ around this point, we recover all the exponents that are 
given in (\ref{expofive}).

\subsection{Proposal for gauge theory effective potentials}

We now wish to use the results of the previous subsections to propose
an effective potential describing the gauge theory. 
While it might be possible to construct such a 
potential in complete generality, in this subsection we study the simplest 
case. We consider the four dimensioal finite temperature gauge 
theory where only one chemical potential is turned on. 

The right order parameter to work with, at the boundary, is the R-charge 
density $\rho_1$. The conjugate chemical potential is $\mu_1$. 
We can now use the first equation of (\ref{thermo3}) to express 
(\ref{poten}) in terms of $\rho_1$. This leads to the following
equation for $\cal F$:
\begin{eqnarray}
{\cal{F}} = &&- \mu_1 \rho_1 - \frac{6 \pi \rho_1}{N_c^2 \mu_1^2} {\sqrt{4 \pi^2 
\rho_1^2 -  
N_c^2\mu_1^3\rho_1}}
+ \frac{12 \pi^2 \rho_1^2}{N_c^2 \mu_1^2}\nonumber\\
&&- \frac{2 \pi \rho_1}{\mu_1} T 
{\sqrt{ - 2\mu_1^2 + \frac{16 \pi^2\rho_1}{\mu_1 N_c^2} - \frac{8 \pi}{\mu_1 N_c^2}
{\sqrt{4 \pi^2 \rho_1^2 - N_c^2 \mu_1^3 \rho_1}} }}.
\label{gaugepot}
\end{eqnarray}
In writing this down, we have reinstated $G$ the five dimensioal 
gravitational constant, various 
factors of $l$ and also used the relation $G = \pi l^3/2N_c^2$. Here 
$N_c$ is the number of colours. 

One can now check that, for  $T > T_c = \pi/\mu_{1c}$, the minimum of 
$\cal{F}$ occurs at
\begin{equation}
\rho_1 = \Big(\frac{N_c^2\mu_1}{32 \pi^2}\Big) 
\frac{\{\mu_1^2 + 2 ( \pi^2 T^2 + {\sqrt{\pi^4 T^4 - \mu_1^2 \pi^2 
T^2}})\}^2}
{ -\mu_1^2 + 2 ( \pi^2 T^2 + {\sqrt{\pi^4 T^4 - \mu_1^2 \pi^2 T^2}})}.
\end{equation}
Now expanding (\ref{gaugepot}) near $T = T_c$, for fixed $\mu_1$, we get
\begin{equation}
\rho_1 - \rho_{1c} \sim (T - T_c)^{\frac{1}{2}},
\end{equation}
where the critical charge density is given by $ \rho_{1c} = 9 \mu_{1c}^3 
N_c^2/(32 \pi^2)$. The above equation leads to $\beta = 1/2$. Similarly, 
one can find the 
other critical exponents given in (\ref{expofive}).

To this end, we would like to make a couple of comments. Firstly, one may 
wish to expand $\cal F$ near $T = T_c$ in powers of $(\rho - 
\rho_c)/\rho_c$ to get a Ginzburg-Landau potential as an expansion in terms of order 
parameter. However, in doing so one does not reproduce the right 
exponents. Higher and higher powers seem to conspire in order to recover 
the results of (\ref{expofive}). Secondly, since the gauge theory is a 
strongly coupled, 
directly computing the potential (\ref{gaugepot}) seems a 
difficult task. However, in \cite{Gubser:1998jb}, an attempt was made to 
construct Ginzburg-Landau potential for the system via a regulated field 
theory model. Unfortunately, all the critical exponents predicted by 
supergravity configuration could not be recovered.

We end this section with a note that, at least for single charge holes, 
similar gauge potentials can be easily constructed in three and six 
dimensions simply using (\ref{lgfour}) and (\ref{lgseven}) along with the 
thermodynamic relations listed in the previous section.

\section{R-charge Conductivity}

So far we have discussed equilibrium thermodynamics of the black holes 
and their dual gauge theories. In this section we will turn towards 
non-equilibrium thermodynamics of this system; in particular we will 
compute the R-charge diffusion constant using gauge-gravity duality. 
There are two ways to compute R-charge diffusion constant. One can 
consider retarded correlation function of R-current and look for a pole  
in the $q^2$-plane where $q$ represents the momentum. Or one can make use 
of Green- Kubo formula to calculate the R-charge conductivity $\lambda$ 
and find R-charge diffusion coefficient $D$ from the relation $\lambda = 
D~.~\chi$, where $\chi$ is the susceptibility matrix. If the 
susceptibility matrix is already known, the advantage with the the second 
method is one need to compute the correlator only  at the zero momentum 
limit and upto first order in frequency. Since, in the present text we 
have already evaluated the susceptibility matrix for various cases we 
will consider the latter method only.

The procedure of computing R-charge diffusion constant using Kubo formula is explained in detail in \cite{Son:2002sd,Policastro:2002se}. For convenience of the reader we recount the essential steps involved in the following. 
For this method we need to find out the retarded correlator
of the R-current:
\begin{equation}
{\tilde G}^{(R)}_{xx}(\omega, \vec{q}) = -i 
\int\limits_{-\infty}^{\infty} ~ dt ~e^{i\omega t}~\theta(t)~
\int ~d\vec{x}e^{-i\vec{q}.\vec{x}}
~ \Big\langle [ J_x(t, \vec{x}) ,J_x(0, \vec{0})]\Big\rangle ~.
\end{equation}
at the zero momentum limit.
The R-charge conductivity $\lambda$ is given in terms of retarded correlator by Green-Kubo formula:
\begin{equation}
\begin{split}
\lambda &= - \lim\limits_{\omega\to 0} 
\frac{ \Im [ {\tilde G}^{(R)}_{xx}(\omega, {\tilde q}=0 )]}{\omega} \nonumber\\
&=  \lim\limits_{\omega\to 0} ~\frac{1}{2\omega}~ 
\int\limits_{-\infty}^{\infty} ~ dt ~ e^{i\omega t} ~ \int ~ d\vec{x} ~\Big\langle [ J_x(t, \vec{x}) ,J_x(0, \vec{0})]\Big\rangle ~.
\end{split}
\end{equation}
The retarded correlator can be computed as follows: We find perturbations in various modes around the black hole solution that satisfy linearized equation following from equations of motion and the `incoming wave boundary condition' at the horizon. Then we calculate the boundary action which can be obtained by plugging in the solutions in the expression of total action evaluated at boundary. The total action consists of three pieces: the bulk action, the Gibbons-Hawking term and the counterterm required to cancel the divergences. The retarded correlator can be obtained by taking derivative of the boundary action with respect to boundary value (of the perturbation in the gauge field) twice. 

Let us demonstarte it in the case of five dimensional black hole with three charges. For the present purpose it is sufficient to consider perturbations in the tensor (metric)  and the vector (gauge fields) modes around the black hole solution and keep the scalars unperturbed. So perturbations are of the form: 
\begin{equation}
g_{\mu\nu}= {\bf g}^{(0)}_{\mu\nu} + h_{\mu\nu}~,
\quad\quad
A^{i}_\mu = {\bf A}^{i(0)}_\mu + {\cal A}^i_\mu~,
\quad i = 1,2\quad.
\end{equation}
In order to determine R-charge conductivity it is enough to  consider perturbations in $(tx)$ and 
$(xx)$ component of the metric tensor and $x$ component of the gauge fields. Moreover one can 
choose the perturbations to depend on radial coordinate $r$, time $t$ and one of the spatial 
worldvolume coordinate $z$.

A convenient ansatz with the above restrictions in mind is
\footnote{Here is a comment about notation in this section. To facillitate comparison with expressions in existing literature we have made a few changes in our notation. In what follows, instead of $r$ we use $u= (r_+/r)^2$ so the boundary is at $u=0$ and horizon is at $u=1$. Similarly the charge parameters $q_i$ will be traded for $k_i =(q_i/r_+^2)$ and we will also introduce $T_0=(r_+/\pi)$.
In this notation the solution for black hole is given by $H_i = (1 + k_i u)$, ${\cal H}= \prod\limits_{i=1}^m H_i$ where $m$ is number of charges. For five dimensional black hole $f= {\cal H} - \prod\limits_{i=1}^m(1+k_i)u^2$ and for four or sweven dimensional black hole $f= {\cal H} - \prod\limits_{i=1}^m(1+k_i)u^3$. We will also use $\tilde\omega = \omega/(2\pi T_0)$}: 
\begin{equation}
h_{tx} = {\bf g}_{0xx}~T(u)~e^{-i\omega t + i q z},\quad
h_{zx} = {\bf g}_{0xx}~Z(u)~e^{-i\omega t + i q z},\quad
{\cal A}^i_x = \frac{\mu}{2} ~ A^i(u)~~e^{-i\omega t + i q z}.\quad
\end{equation}
Here $\omega$ and $q$ represent the freequency and momentum in $z$ direction respectively and we set perturbations in the other components to be equal to zero.

Next step is to find linearized equations following from the equations of motion at zero momentum limit that the perturbations has to satisfy. It turns out that in linearized equation at zero momentum limit metric perturbation $Z(u)$ decouple from the rest. One can further eliminate $T(u)$ reducing it to equation for perturbations in gauge field only. For five dimensional black hole with all the three charges the equation becomes in this notation
\begin{equation}
\begin{split}
&A_1^{\prime\prime} + ~\left(
\dfrac{f^\prime}{f} + \dfrac{H_1^\prime}{H_1} -  \dfrac{H_2^\prime}{H_2} -
 \dfrac{H_3^\prime}{H_3} \right) A_1^\prime
 + \dfrac{\tilde\omega^2 {\cal H}}{f^2 u} A^1 
- \dfrac{u(1+k_1)}{fH_1^2}
\Big[ k_1(1+k_2)(1+k_3) A_1 \\&
+ k_2(1+k_1)(1+k_3)A_2 +  k_3(1+k_1)(1+k_2)A_3 \Big] = 0, \label{D3eq}
\end{split}
\end{equation}
while equation for the other two components can be obtained 
by cyclically permuting $(123)$.

We require solutions to these equations with ``incoming wave boundary condition'' at the horizon at $u=1$.  Since $f(u)$ vanishes at $u=1$ one can solve the indicial equation near $u=1$ and obtain two solutions which have asymptotic behaviour near $u=1$ as $(u-1)^\nu$ with $\nu = \pm i{\tilde\omega}(T_0/2T)$. The incoming wave function corresponds to the negative value and so incorporating the boundary condition corresponds to the following ansatz:
\begin{equation}
A_i = \dfrac{ f^{-i\tilde\omega (T_0/2T)}}{1 + k_i u} a_i(u),
\quad\quad i = 1,2,3.\label{D3ansatz}
\end{equation} Since in the sequel we need solutions upto first order of $\omega$ only we can write it as
\begin{equation}
a_i(u) = p_i(u) + i\omega s_i(u) + 0(\omega^2).\label{aiu}
\end{equation}
In terms of $a_i$'s the boundary action appears to be
\begin{equation}
\text{S}_{\text{boundary}} = \lim_{u\to 0} \dfrac{N_c^2 T_0^2}{16}
~\int ~ dtd\vec{x}~ \dfrac{f}{\cal H} 
[ H_1^2 a_1^\prime a_1 + H_2^2 a_2^\prime a_2 + H_3^2 a_3^\prime a_3],
\end{equation}
where we have kept terms quadratic in perturbation. The retarded correlator can be obtained by taking derivative of the total action with respect to boundary value twice. 

Proceeding further with this generic charge leads to expressions, which are technically complicated. Moreover, 
one unusual problem here is that we cannot obtain special cases, such as two charge solution, 
starting from generic cases (three charge) by setting one of the charges to zero. This happens 
because obtaining these equations involves rescaling of fields $A^i(u) $ with respective charges 
and forces us to do the analysis in each case separately. Therefore, we will restrict ourselves to 
the R-charge conductivity (for black holes in five, four and seven dimensions) in the simplest 
cases only, deferring analysis of more complicated cases to the appendix. We will also adjoin a 
brief outline of the general method that has been used to solve all these cases.


\subsection{Five dimensional black hole with two charges}

In this case we have two $U(1)$ gauge fields. The solution is
\begin{equation}
H_1 = (1+k_1u),\quad
H_2 = (1+k_2u),\quad
{\cal H} = H_1 H_2\quad
f(u) = {\cal H} - (1+k_1)(1+k_2)u^2.
\end{equation} 
The perturbed equations are given by,
\begin{equation}
\begin{split}
&A_1^{\prime\prime} + \left( \dfrac{f'}{f} - \dfrac{\cal H^\prime}{\cal H} + 2 
\dfrac{H_1^\prime}{H_1} \right) A_1^\prime + 
\dfrac{{\tilde\omega}^2 {\cal H}} {u f^2}A_1 \\
&- \dfrac{u (1+k_1)}{f H_1^2} (k_1 (1+k_2)A_1+ k_2(1+k_1)A_2) 
= 0,\label{A1}
\end{split}
\end{equation} 
\begin{equation}
\begin{split}
&A_2^{\prime\prime} + \left( \dfrac{f'}{f} - \dfrac{\cal H^\prime}{\cal H} + 2 
\dfrac{H_2^\prime}{H_2} \right) A_2^\prime + 
\dfrac{{\tilde\omega}^2 {\cal H}} {u f^2}A_2 \\
&- \dfrac{u (1+k_2)}{f H_2^2} (k_1 (1+k_2)A_1+ k_2(1+k_1)A_2) 
= 0.\label{A2}
\end{split}
\end{equation}
. We observe that the equation for $A_2$ can be obtained from equation for $A_1$ by interchanging $k_1,H_1, A_1$ with $k_2,H_2, A_2$ respectively.
From now on for convenience, we will write perturbed equation for only one of the gauge fields. The ansatz is given by
\begin{equation}
A_i = \dfrac{ f^{-i\tilde\omega (T_0/2T)}}{1 + k_i u} a_i(u),
\quad\quad i = 1,2,3.\label{D3ansatzone}
\end{equation}
We need to solve for $a_i(u)$ in terms of the boundary value of the gauge fields. Since in the sequel we will need
solutions upto first order of $\omega$ only we can write it as
\begin{equation}
a_i(u) =  \left[ p_i(u) + i\tilde\omega s_1(u) + 0({\tilde\omega}^2) \right].
\end{equation}
where $i$ takes value upto 2 for this particular case.
It turns out from (\ref{A1}), that $p_1$ and $s_1$ satisfy the following equations respectively:
\begin{equation}
\begin{split}
&p_1^{\prime\prime}(u) + \left( \dfrac{f^\prime}{f}-
\dfrac{\cal H^\prime}{\cal H}  
\right) p_1^\prime (u)\nonumber \\
& -
\dfrac{k_1}{H_1}  \left(\dfrac{f^\prime}{f}-
\dfrac{\cal H^\prime}{\cal H}  
\right) p_1(u) - u \dfrac{1+k_1}{f H_1} \left[\dfrac{k_1 (1+k_2)}{H_1}p_1+\dfrac{k_2(1+k_1)}{H_2}p_2\right] 
= 0,
\end{split}
\end{equation}
\begin{equation}
\begin{split}
& s_1^{\prime\prime}(u) + \left( \dfrac{f^\prime}{f}-
\dfrac{\cal H^\prime}{\cal H}
\right) s_1^\prime (u)- \dfrac{T_0}{T}\dfrac{f^\prime}{f} p_1^\prime (u)+\dfrac{T_0}{2 T}\left(\dfrac{f^\prime \cal H^\prime}{f \cal H}-
\dfrac{f^{\prime\prime}}{f}\right) p_1 +
\nonumber\\
&\dfrac{k_1}{H_1}  \left(\dfrac{f^\prime}{f}-
\dfrac{\cal H^\prime}{\cal H}  
\right) s_1(u)-u \dfrac{1+k_1}{f H_1} \left[\dfrac{k_1 (1+k_2)}{H_1}s_1+\dfrac{k_2(1+k_1)}{H_2}s_2\right]
 =0.
\end{split}
\end{equation}
Similarly by interchanging $k_1$, $p_1$, $s_1$ with $k_2$, $p_2$, $s_2$ 
we obtain equation for $p_2$ and $s_2$. According to our boundary condition 
all these solutions need to be regular at $u=1$. Without any loss of generality we can choose $s_i =0$ .

The solution for $p_i(u)$ are given below:
\begin{equation}
\begin{split}
p_1(u) &= b_{10} ( 1 + \dfrac{k_1 u}{2} ) - 
b_{20} \dfrac{k_2}{2} \dfrac{1+k_1}{1+k_2} u,
\\
p_2(u) &= b_{10} ( 1 + \dfrac{k_2 u}{2} ) - 
b_{20} \dfrac{k_1}{2} \dfrac{1+k_2}{1+k_1} u,
\end{split}
\end{equation}
where $b_{10}$ and $b_{20}$ are boundary value of $A_1$ and $A_2$ respectively.
The solutions of $s_i(u)$ are more involved and given by
\begin{eqnarray}
s_1(u) &= &h_1 u + (h_2 + h_3 u) \log{\left[1 + (1+k_1+k_2)u\right]},
\nonumber\\
s_2(u) &=& l_1 u + (l_2 + l_3 u) \log{\left[1 + (1+k_1+k_2)u\right]}.
\end{eqnarray}
Note that at $u =0$, $s_i(u)=0$ as mentioned before.
The constants are given by 
\begin{equation}
h_1 =  b_{10} J_{10} + b_{20} J_{20},
\quad\quad
h_2 =  b_{10} L_{10} + b_{20} L_{20},
\end{equation}
where
\begin{equation}\begin{split}
J_{10} &= k_1 \left[\dfrac{
(k_1^3 - k_1^2 (k_2-3) + k_2 (k_2^2 - k_2 -6) + k_2 (k_2^2 + 3 k_2 + 2)
}{ 4(1+k_1)(1+k_2)(1+k_1+k_2)}
\right] ,
\nonumber\\
J_{20} &= -  k_2 \left[\dfrac{
k_1^3 - k_1^2(k_2 - 3) - k_1(k_2^2 + 2 k_2 -2) + k_2 ( k_2^2 + 3 k_2 + 2)
}{ 4 ( 1 + k_2)^2 (1 + k_1 + k_2)}   \right].
\end{split}
\end{equation}
The values of $L_{10}$ and $L_{20}$ are
\begin{equation}
\begin{split}
L_{10} = & \left[\dfrac
{2+2k_1^2+2k_2+k_2^2+k_1(4+3k_2)}{(1+k_1+k_2)^2}\right],\\
L_{20}= & - \left[ \dfrac{(1+k_1)k_2(2+k_1+k_2)}{(1+k_2)(1+k_1+k_2)^2} \right].
\end{split}
\end{equation}
$h_3$ is given by
\begin{equation}
h_3 = b_{10} k_1 - \dfrac{k_2(1+k_1)}{1+k_2}b_{20}.
\end{equation}
$l_1$, $l_2$ and $l_3$ are given by
\begin{equation}
l_1 =  b_{10} {\tilde J}_{10} + b_{20} {\tilde J}_{20},
\quad\quad
l_2 =  b_{10} {\tilde L}_{10} + b_{20} {\tilde L}_{20},
\end{equation}
where ${\tilde J}_{10}$(${\tilde J}_{20}$) can be obtained by interchanging $k_1$ with $k_2$ 
in $J_{20}$($J_{10}$) and 
 ${\tilde L}_{10}$(${\tilde L}_{20}$) can be obtained by interchanging $k_1$ with $k_2$ in $L_{20}$($L_{10}$)

The second step consists of evaluation of the boundary action 
quadratic in the boundary values. The boundary action turns out
to be of the following form. Since we need only imaginary contribution of 
$0({\tilde\omega}$) we keep the other terms implicit.
\begin{equation}
\begin{split}
\text{S}_{\text{boundary}} =& \lim_{u\to 0} \dfrac{N_c^2 T_0^2}{16}
~\int ~ dtd\vec{x}~ \dfrac{f}{\cal H} 
\left[ H_1^2 a_1^\prime a_1 + H_2^2 a_2^\prime a_2 \right]
\nonumber\\
=&  {\cal {O}}(\tilde\omega^0) + \dfrac{N_c^2 T_0^2}{16} 
(-i{\tilde\omega}) \bigg[
b_{10}^2 \left[ J_{10} + {\tilde J}_{10} (1+k_1+k_2) - (T_0/2T)(k_1+k_2)\right]
\nonumber\\
+& b_{10} b_{20} \left[J_{20} {\tilde L}_{20} + 
(1+k_1+k_2)({\tilde J}_{20} + {\tilde L}_{20}) \right]
\nonumber\\
+& b_{20}^2 \left[ L_{10} + {\tilde L}_{10}(1+k_1+k_2) - (T_0/2T)(k_1+k_2)\right]
\bigg] + {\cal{O}}({\tilde\omega}^2)
\end{split}
\end{equation}

From the above boundary action we obtain the following components of the retarded Green's function
\begin{equation}
\begin{split}
{\tilde G}^{(RR)}_{11}({\tilde\omega}, \vec{q}=0) 
=& {\cal {O}}(\tilde\omega^0) -i{\tilde\omega}\dfrac{N_c^2 T_0^2}{8}
\left[ J_{10} + {\tilde J}_{10} (1+k_1+k_2) - (T_0/2T)(k_1+k_2)\right] + {\cal{O}}({\tilde\omega}^2),
\\
{\tilde G}^{(RR)}_{12}({\tilde\omega}, \vec{q}=0)  =& {\cal {O}}(\tilde\omega^0) - 
i{\tilde\omega}  \dfrac{N_c^2 T_0^2}{16} 
\left[ J_{20} {\tilde L}_{20} + 
(1+k_1+k_2)({\tilde J}_{20} + {\tilde L}_{20}) \right] + {\cal{O}}({\tilde\omega}^2), \\
{\tilde G}^{(RR)}_{22}({\tilde\omega}, \vec{q}=0)  =&  {\cal {O}}(\tilde\omega^0) - 
i{\tilde\omega} \dfrac{N_c^2 T_0^2}{8} 
\left[ L_{10} + {\tilde L}_{10}(1+k_1+k_2) - (T_0/2T)(k_1+k_2)\right] + {\cal{O}}({\tilde\omega}^2).
\end{split}
\end{equation}

The components of the R-charge conductivity then obtained as
\begin{equation}
\begin{split}
\lambda_{11}
=& 
2. \dfrac{N_c^2 T_0^2}{16}
\left[ J_{10} + {\tilde J}_{10} (1+k_1+k_2) - (T_0/2T)(k_1+k_2)\right] 
\\
\lambda_{12} =& 
\dfrac{N_c^2 T_0^2}{16} 
\left[ J_{20} {\tilde L}_{20} + 
(1+k_1+k_2)({\tilde J}_{20} + {\tilde L}_{20}) \right] \\
\lambda_{22} =& 2. \dfrac{N_c^2 T_0^2}{16} 
\left[ L_{10} + {\tilde L}_{10}(1+k_1+k_2) - (T_0/2T)(k_1+k_2)\right].
\end{split}
\end{equation}

The susceptibility matrix $\chi_{ij}$ can be written in this
notation as
\begin{eqnarray}
\chi_{ij} = \dfrac{{\bar r}^2}{16\pi}
\left(
\begin{array}{cc}
k_1^2 + k_1k_2 + k_2 -5 k_1 -2 & \sqrt{k_1k_2} (k_1 + k_2)\\
 \sqrt{k_1k_2} (k_1 + k_2) & k_2^2 + k_2k_1 + k_1 -5 k_2 -2
\end{array} 
\right) \dfrac{1}{k_1+k_2-2}
\end{eqnarray}
The components of the diffusion coefficient can be
obtained by using the relation ${\cal D}=\lambda.\chi^{-1}$. Since
all the components of susceptibility diverges on the critical line
$k_1 + k_2 =2$ and conductivity remains finite it follows that 
all the components of the diffusion coeficient vanish. Vanishing 
of the diffusion coefficients is consistent with slowing down at
critical point.

\subsection{Four dimensional black holes}

Next we consider the black holes in three dimensions with two charges (single charged case has 
already been discussed in \cite{Maeda:2008hn}).
There can be maximum four charges 
as there are four commuting $U(1)$s. Leaving the more general cases for the appendix, here we 
consider black holes with two $U(1)$s. The equation for gauge field is given by\begin{equation}
\begin{split}
A_1^{\prime\prime} &+ \left( \dfrac{f'}{f} - \dfrac{\cal H^\prime}{\cal H} + 2 
\dfrac{H_1^\prime}{H_1} \right) A_1^\prime + 
\dfrac{{\tilde\omega}^2 {\cal H}} {u f^2}A_1 \\
&- \dfrac{u^2 (1+k_1)}{f H_1^2} \left(k_1 (1+k_2)A_1+ k_2(1+k_1)A_2\right) 
= 0. \label{M2eq}
\end{split}
\end{equation}
Equation for $A_2$ can be obtained by inchanging $(A_1, k_1, H_1)$ with $(A_2, k_2 , H_2)$ respectively. The ansatz consistent with the boundary condition is 
\begin{equation}
A^i(u) = \dfrac{f^{-i{\tilde\omega}(T_0/2T)}}{1 + k_i u} a_i(u),
\quad\quad
a_i(u) = p_i(u) + i\tilde\omega s_i(u). \label{M2ansatz}
\end{equation}
Using above ansatz we obtain equations for $p_1,p_2,s_1,s_2$.
The solution for $p_1$ is given by,
\begin{equation}
p_1(u) = b_{10}(1 + \frac{2k_1}{3} u ) + b_{20} \dfrac{k_2(1+k_1)}{3(1+k_2)}u,
\end{equation} 
while $p_2$ can be obtained from above expression by permuting $k_1$ and $k_2$.
The equations for $s_i$ are more involved and can be written as
\begin{equation}
\begin{split}
&s_1(u)^{\prime\prime} + s_1^\prime (\dfrac{f^\prime}{f} - 
\dfrac{{\cal H}^\prime}{\cal H}) -  s_1(u)\dfrac{k_1}{H_1}(\dfrac{f^\prime}{f} - \dfrac{{\cal 
H}^\prime}{\cal H}) \\ & -  \dfrac{(1+k_1)u^2}{H_1}\left[ \dfrac{k_1(1+k_2)}{f H_1}s_1(u) - 
\dfrac{k_2(1+k_1)}{f H_2}s_2(u)\right]\\ &- 2 v p_1(u)^\prime \dfrac{f^\prime}{f}+ v 
p_1(u)\left( \dfrac{f^\prime}{f}\dfrac{{\cal H}^\prime}{\cal H}- \dfrac{f^{\prime\prime}}{f}\right) = 0 ,
\end{split}
\end{equation}
where we used $v=T_0/(2T)$. Equation for $s_2$ can be obtained by permutation, as usual. Solution is given by,
\begin{equation}
\begin{split}
s_1(u) = &h_1 u + (h_2 + h_3 u) \arctan\left(\dfrac{\sqrt{ 3 - k_1^2 + 2k_2 - k_2^2 +2k_1(1+k_2)}u}{2 + (1+k_1+k_2)u}\right)\\ + &(h_4 + h_5 u) \log(\dfrac{f}{1-u}),
\end{split}\end{equation}
and similarly for $s_2(u)$.
The parameters are as follows:
\begin{eqnarray}
h_1 &=& \bigg(
(3+2k_1+2k_2+2k_1k_2)
\Big[
-2(1+k_1)k_2(k_1^3+k_1^2(1-k_2)-k_1k_2(1+k_2)+k_2^3(1+k_2))
 \Big]b_{20}\nonumber\\ &+& 
 k_1(1+k_2)\Big[4k_1^3-k_1^2(7k_2-4)+(k_2+1)k_2^2+k_1k_2(2k_2-7)\Big] b_{10} \bigg)\nonumber\\
&\times& \dfrac{v}{9(1+k_1)(1+k_2)^2[-3+k_1^2-2k_2+k_2^2-2k_1-2k_1k_2]},\\
h_2 &=& \dfrac{2h_3}{k_1} +\dfrac{(1+k_1)}{k_1(1+k_2)}l_3, \\
h_3 &=& \bigg( 2k_1(1+k_2)\Big[9+k_1^3++3k_2+k_2^2+k_2^3+k_1^2(5k_2+13)+ k_1(21 + 8k_2 - k_2^2)\Big] b_{10}\nonumber\\& - & k_2(1+k_1)\Big[9 + k_1^3 + 21 k_2 + 13 k_2^2 + k_2^3 + k_1^2(13 + 5k_2)+ k_1(21 + 20 k_2 + 5 k_2^2)\Big]b_{20}\bigg)\nonumber\\
&&\dfrac{v}{3(1+k_2)(3-k_1^2+2k_2-k_2^2 + 2k_1 + 2k_1k_2)^{3/2}},\\
h_4 &=& (3/2) v b_{10},\\
h_5 &=& k_1 v b_{10} - k_2\dfrac{1+k_1}{2(1+k_2)}v b_{10},
\end{eqnarray}
where we have used $v=(T_0/2T)$ and the parameter $l_3$:
\begin{equation}
\begin{split}
l_3 &= \bigg(2(1+k_1)k_2
\left[ 9 + k_1^3 - k_1^2(k_2-1) + 21 k_2 + 13 k_2^2 + k_2^3 + 
k_1(3+ 8k_2 + 5 k_2^2) \right] b_{20} 
\\&
- k_1(1+k_2)\left[ 9 + k_1^3 + 21 k_2 + 13 k_2^2 + k_2^3 + 
k_1^2(13 + 5k_2) + 21 k_1 + 20 k_2 + 5 k_2^2 \right]b_{10}\bigg)
\\& 
\dfrac{v}{3(1+k_1)(3-k_1^2+2k_2-k_2^2+ 2k_1(1+k_2)^{3/2}}
\end{split}
\end{equation}
The boundary action is given by
\begin{equation}
\text{S}_{\text{boundary}} = \lim_{u\to 0} 
\dfrac{
N_c^{\frac{3}{2}} ~T_0
}
{36\sqrt{2}}
~\int ~ dtd\vec{x}~ \dfrac{f}{\cal H} 
\left[ H_1^2 a_1^\prime a_1 + H_2^2 a_2^\prime a_2 \right].
\end{equation}
Now we can follow the same procedure as in the $d=5$ case and obtain
the R-charge conductivity by differentiating twice with respect to
$b_{10}$ and $b_{20}$. The $(11)$ and $(12)$ components of R-charge conductivity turn out to be
\begin{equation}
\begin{split}
\lambda_{11} &=\dfrac{N_c^{\frac{3}{2}} ~T_0 v}{36\sqrt{2}} 
\dfrac{2 \left[ 4k_1^3(2+k_2) + 9 ( 3 + 3k_2 + k_2^2) + k_1^2(45 + 31 k_2 + k_2^3) + k_1(63 + 54 k_2 + 11 k_2^2)\right] }{9(1+k_1)(1+k_2)},\\
\lambda_{12} &= \dfrac{
N_c^{\frac{3}{2}} ~T_0 v}{36\sqrt{2}}
\dfrac{2}{9(1+k_1)^2(1+k_2)^2} \Big[ k_1^4k_2(2+k_2) + k_2(9+9k_2+2k_2^3) + k_1^3(2+18k_2+ 13 k_2^2 + 2k_2^3) + \\&
k_1^2(9+54 k_2 + 48k_2^2 +13 k_2^3 + k_2^4) + k_1(9+54 k_2 + 54 k_2^2 + 18 k_2^3 + 2k_2^4)\Big].
\end{split}
\end{equation}
The $(22)$ component can be ontained by interchanging $k_1$, $b_{10}$ with $k_2$, $b_{20}$ respectively.
Once again since all the components of susceptibility diverges near critical point, while conductivity remains finite, as evident from the above equations, the diffusion coefficients will vanish.
 
\subsection{Seven dimensional black hole with two charges}

The seven dimensional case corresponds to the M5 brane. Since the transverse 
space is five dimensional only two $U(1)$ charges are possible. The equations
for the perturbation in gauge field is given by
\begin{equation}
A_1^{\prime\prime} + \left( \dfrac{f^\prime}{f} - 
\dfrac{{\cal H}^\prime}{\cal H} + 2 \dfrac{H_1^\prime}{H_1} - \dfrac{1}{u}\right) A_1^\prime - \dfrac{4(1+k_1) u^3}{f H_1^2} \Big[ k_1(1+k_2) A_1 + k_2(1+k_1)A_2 \Big] = 0 . \end{equation}
The other equation can be obtained by interchanging $A_1$, $k_1$ with $A_2$ and $k_2$ respectively.
The ansatz appropriate to the boundary condition is
\begin{equation}
A_1(u) = \dfrac{f^{-i\tilde\omega v}}{H_1} \left(p_1(u) + i\tilde\omega s_1(u)\right).
\end{equation}
The explicit form of the solution to $s_i(u)$'s are obtained but they are not written here as they are quite complicated. But the
conclusion remains the same i.e. as we approach critical line diffusion constant goes to zero.
We explain in the appendix (see last section) why two-charge case for 
M5 brane is complicated than the same for D3 and M2 brane. 


\section{Conclusion}

 In this paper we have investigated some of the features of equilibrium and non-equilibrium 
 thermodynamic properties of R-charged AdS black hole in its full generality near the critical 
point. Furthermore, by making use of the gauge-gravity correspondence, we tried to extract 
some properties of strongly coupled boundary gauge theories at finite temperature. More 
specifically, we analyzed five, four and seven dimensional black holes with multiple charges. 
This led to the construction of Bragg-Williams potentials associated with these black holes.
We discussed how Bragg-Williams potential allow us to make a proposal for the gauge 
theory effective potential (above the critical temperature). This effective potential is 
usefull in determining the equilibrium properties of the strongly coupled gauge theories in 
the presence of non-zero chemical potential. Subsequently, we analysed some of the 
non-equilibrium properties of the black holes and its duals. Though, for single R-charged 
black holes, 
thermal properties were discussed in \cite{Son:2006em} and \cite{Maeda:2008hn}, 
we extended these results in several ways. Besides providing a general framework to solve the 
perturbed gravity equations, we determined the conductivities and diffusion coefficients for
multiply charged black holes. 

Our work may be extended futher by considereing rotating R-charged black holes. They were 
consuructed, for example, in \cite{Chong:2006zx}. Another question that is of crucial importance 
is to understand the fate of these R-charged black holes below the critical temperature. As we 
saw, these holes become unstable. However, the new phase to which it crosses over is, 
to our knowledge, not yet clearly understood. 
   
\section*{Acknowledgements}

We have benefitted from useful conversations and communications 
with Somen Bhattacharjee, Sankhadeep Chakrabarty, Sumit Das, Alok Kumar, Kengo Maeda, Makoto Natsuume, Takashi 
Okamura, Binata Panda, Kalyana Rama, Bala Sathiapalan, Sreekumar Sengupta, and Goutam Tripathy. We thank organizers and 
participants of ISM 08 at Pondicherry where part of this work was presented.

\renewcommand{\thesection}{\Alph{section}}
\setcounter{section}{0}
\section{General solution to linearized equation for perturbation}
In this appendix we briefly sketch the structure of the solutions and the general method that have been used to solve linearized equations for the perturbations in the gauge field. The equations satisfied by $p_i$ and $s_i$ in each of the cases, {\it i.e.} for five four and seven dimensional black hole can be written in the following general form:
\begin{equation}
\begin{split}
&p_i^{\prime\prime}(u) + \left( \dfrac{f^\prime}{f}-
\dfrac{\cal H^\prime}{\cal H}+\dfrac{c}{u}  
\right) p_i^\prime (u)\\
&-
\dfrac{k_i}{H_i}  \left(\dfrac{f^\prime}{f}-
\dfrac{\cal H^\prime}{\cal H}  
\right) p_i(u)- u^{n} \dfrac{1+k_i}{f H_i} ~\prod\limits_{i=1}^m (1+k_i)(\sum\limits_{i=1}^m \dfrac{k_i}{(1+k_i)H_i} p_i ) 
= 0,\quad i=1,..,m.
\end{split}
\end{equation}
\begin{equation}
\begin{split}
& s_i^{\prime\prime}(u) + \left( \dfrac{f^\prime}{f}-
\dfrac{\cal H^\prime}{\cal H}+\dfrac{c}{u}
\right) s_i^\prime (u)- \dfrac{T_0}{T}\dfrac{f^\prime}{f} p_i^\prime (u)+\dfrac{T_0}{2 T}\left(\dfrac{f^\prime \cal H^\prime}{f \cal H}-
\dfrac{f^{\prime\prime}}{f}\right) p_i 
\\
&-\dfrac{k_i}{H_i}  \left(\dfrac{f^\prime}{f}-
\dfrac{\cal H^\prime}{\cal H}  
\right) s_i(u) -  u^{n} \dfrac{b(1+k_i)}{f H_i} ~\prod\limits_{i=1}^m (1+k_i)(\sum\limits_{i=1}^m \dfrac{k_i}{(1+k_i)H_i} p_i)
 =0,\quad i=1,..,m,
\end{split}
\end{equation}
where $m$ is the number of R-charges.
The parameters $c$, $n$, $b$ and $m$ have following values for the different cases:
\begin{itemize}
\item Five dimensional black hole: $c=0$, $n=1$, $b=1$, $m=1,2,3$.
\item Four dimensional black hole: $c=0$, $n=2$, $b=1$, $m=1,2,3,4$.
\item Seven dimensional black hole: $c=-1$, $n=3$, $b=4$, $m=1,2$.
\end{itemize}
The solutions should be regular at horizon {\it i.e.} at $u=1$. As already mentioned in the main text, without any loss of generality we can also impose the condition that at $u=0$ the term linear in frequency should vanish.

Solution for $p_i$'s can be written generically as
\begin{equation}
p_i(u) = b_{i0} + c_{i0}~ u^{l},\label{generalp}
\end{equation}
where $c_{i0}$ depends on all the $k_i$'s and $b_{i0}$'s. 
Clearly $b_{i0}$ is the constant representing boundary value of
$A_i(u)$. For five and four dimensional black hole $l=1$ while 
for seven dimensional black hole $l=2$.

Now we turn to solution for $s_i$'s.
The function $f(u)$ in the metric generically can be written as, 
\begin{equation}
f=(1-u)F(u),
\end{equation}
where $F(u)$ is a polynomial in $u$ of order $\nu$, and so has $\nu$ number of zeroes given by $u=u_j, j= 1,2,..,\nu$. For four and five dimensional black hole $\nu=m-1$ while for seven dimensional blackhole $\nu = m+2$

Then most general solution for $s_{i}$ can be written as
\begin{equation}
s_{i}= P_i(u) +\sum\limits_{j=1}^\nu Q_{ij}(u)\log(1-u/u_j),\label{generals}
\end{equation}
where $P_i(u)$ and $Q_{ij}(u)$ are two sets of functions.
In order to ensure $s_i(0) = 0$, we have $P_i(0)=0$. 
We require our solution to be regular at the horizon so $\log(1-u)$ term is absent.  This is the genral form representing the solution for various cases as discussed in main text as well as in appendix. 
For each particular case we need to determine the functions $P_i(u)$ and $Q_{ij}(u)$.

\section{Five dimensional black hole with three charges}

In the case of five dimensional black hole with three charges the values of the parameters are $c=0$, $n=1$, $b=1$, $m=3$ and $l=1$. 
Since $f=(1-u)(1+(1+k_1+k_2+k_3) u - k_1 k_2 k_3 u^{2})$ we get
\begin{equation}
F(u) = (1+(1+k_1+k_2+k_3) u - k_1 k_2 k_3 u^{2}),
\end{equation}
which is a quadratic polynomial implying $\nu=2$. It is helpful for calculation if we write the second term in (\ref{generals}) as
\begin{equation}
Q_{i1} \log(1-u/u_1)+ Q_{i2} \log(1-u/u_2) =\dfrac{Q_{i1}+Q_{i2}}{2}\log(F(u))+\dfrac{Q_{i1} - Q_{i2}}{2}\log\Big(\dfrac{u_2(u_1 -u)}{u_1(u_2 -u)}\Big). 
\label{logcombine}
\end{equation}

Using $l=1$ we write down $p_i(u)$ from (\ref{generalp}) as
\begin{equation}
p_i(u) = b_{i0} + c_{i0}u.
\end{equation}
$c_{10}$ turns out to be
\begin{equation}
c_{10} = \dfrac{1}{2} \left[
 k_1 b_{10} - (1+k_1)\left( \dfrac{k_2}{1+k_2} b_{20}
+  \dfrac{k_3}{1+k_3} b_{30} \right)\right],
\end{equation}
Other components of $c_{i0}$ can be obtained
by cyclic permutation of $(123)$.

Expressions of $s_i(u)$ are as follows:
\begin{equation}
\begin{split}
s_1(u) &= h_1 u + (h_2 + h_3 u) \log \left(
\dfrac{2 + (1+k_1+k_2+k_3 + \sqrt{\triangle})u } 
{2 + (1+k_1+k_2+k_3- \sqrt{\triangle})u }\right)\\
&+ (h_4+h_5u)\log [1 + (1 + k_1 + k_2 +k_3)u - k_1k_2k_3u^2],
\end{split}
\end{equation} 
while
\begin{equation}
\triangle = (1+k_1+k_2+k_3)^2 + 4k_1k_2k_3.
\end{equation}
$h_i$ for $i=1,2,3,4,5$ are constants, which depend on $k_1$, $k_2$ and $k_3$ and
boundary values of the gauge fields, namely, $b_{10}$, $b_{20}$, $b_{30}$. 
Introducing some additional parameters, $h_1$ can be written succintly as
\begin{equation}
h_1 = \dfrac{(2+k_1+k_2+k_3-k_1k_2k_3)v}{4(1+k_2)(1+k_3)\triangle}
\left[ \dfrac{k_1J_1}{1+k_1} b_{10}
-  \dfrac{k_2J_2}{1+k_2} b_{20}
- \dfrac{k_3J_3}{1+k_3} b_{30} \right].
\end{equation}
These new parameters $J_1$, $J_2$ and $J_3$ are fourth order polynomials 
in $k_i$'s and are given as
\begin{equation}
\begin{split}
J_1 = &k_1^4 - k_1^3(-2+k_2+k_3) - k_1^2(k_2-k_3)^2 + 4(k_2+k_3)(k_1^2+k_2k_3) + k_1^2 - 4k_2k_3 \\&+ k_1(k_2^3 - 2k_2^2 -5k_3k_2^2 - 3k_3-2k_3^2+k_3^3-3k_2-5k_2k_3^2),\\
J_2 = &[k_1^3 + k_2^3 - k_3^3 - k_1^2k_2 - k_1k_2^2 + 3k_1^2k_3 +3k_2k_3^2 + 10 k_1k_2k_3 ] \\
&+ [2(-k_1^2 + k_2^2 +6k_3^2) + 4(k_1 + k_2)k_3 ] + [k_1+k_2+7k_3],\\
J_3 =&[ k_1^3 - k_2^3 + k_3^3 - 3k_1^2k_2 + 3k_1k_2^2 + 3k_2^2k_3 - 3k_2k_3^2 -k_1^2k_3-k_1k_3^2 + 10 k_1k_2k_3 ]\\
&+ [2k_1^2 + 6k_2^2 +2k_3^2+4k_2k_3+4k_1k_2]+ [k_3+7k_2+k_1].
\end{split}
\end{equation}
Similarly $h_2$ is given by,
\begin{equation}
\begin{split}
h_2 &= - \dfrac{1+k_1}{2 \sqrt{\triangle}}
\bigg[
2k_1(1+k_2)(2+k_1+k_2+k_3-k_1k_2k_3)
\Big[\dfrac{k_2L_2}{1+k_2}b_{20}+\dfrac{k_3 L_3}{1+k_3}b_{30}\Big]
\\&
+ (1+k_2)(1+k_3)L_1b_{10}\bigg] v ,
\end{split}
\end{equation}
where $L_i$'s are
\begin{equation}
\begin{split}
L_1 &= -1 + k_2^3+k_3 +3k_3^2 + k_3^3 +3k_2^2+3k_2^2k_3-k_1^3+2k_1^3k_2k_3+k_2+6k_2k_3+3k_2k_3^2-3k_1+k_1k_3^2
\\&
+16k_1k_2k_3+6k_1^2k_2k_3^2+k_1k_2^2+6k_1k_2^2k_3+k_1^2[-3-k_3+2k_2^2k_3-k_2+8k_2k_3+2k_2k_3^2],\\
L_2 &=(1+k_1+k_2-k_3),\\
L_3 &=(1+k_1-k_2+k_3).
\end{split}
\end{equation}
$h_3$ is given by
\begin{equation}
h_3 = \dfrac{k_1}{2} h_2 - \dfrac{v}{2(1+k_2)(1+k_3){\triangle^{3/2}}} 
(k_2 M + k_3 N),
\end{equation}
where $v$ depends on $k_i$'s in the following manner
\begin{equation}
v = \dfrac{
\sqrt{(1+k_1)(1+k_2)(1+k_3)}
	}
{2+k_1+k_2+k_3-k_1k_2k_3}.
\end{equation}
$M$ is given as follows.
\begin{equation}
\begin{split}
M=&-2k_2(1+k_2)(2+k_1+k_2+k_3-k_1k_2k_3)\Big[ k_1(1+k_3)(1+k_1+k_2-k_3)b_{10} \\&+ k_3(1+k_1)(1-k_1+k_2+k_3)b_{30} \Big] + \\
&(1+k_1)(1+k_3) 
\Big[ -1 + k_1^3 -k_2^3 + k_3 + 3k_3^2 + k_3^3 -3k_2^2 - k_3k_2^2 \\
&+ k_1^2 ( 3 + k_2 + 3k_3 +6k_2k_3 + 2k_2^2k_3)
+ k_2(-3+k_3^2) + \\ & k_1(1+6k_3+2k_2^2k_3+3k_3^2+16k_2k_3+6k_2k_3^2-k_2^2+8k_3k_2^2+2k_2^2k_3^2)\Big] b_{20}.
\end{split}
\end{equation}
$N$ can be obtained by interchanging $k_2$ and $k_3$ in the above expression. 
The rest of the parameters $h_4$ and $h_5$ are
\begin{equation}
\begin{split}
h_4 &= \dfrac{3v}{2} b_{10}\\
h_5 &= \dfrac{3v}{4}(1+k_1)\left[ \dfrac{k_1}{1+k_1} b_{10} - \dfrac{k_2}{1+k_2}b_{20} - \dfrac{k_3}{1+k_3}b_{30} \right] (1+k_1) 
\end{split}
\end{equation}
Evaluating the boundary  action
\begin{equation}
S_{\text{boundary}} = \dfrac{N_c^{(3/2)}T_0}{36\sqrt{2}}~
\lim\limits_{u\to 0}~ \int dt d\vec{x} ~ \dfrac{f}{\cal H}
\left(H_1^2 a_1^\prime a_1 + H_2^2 a_2 a_2^\prime + H_3^2 a_3 a_3^\prime \right),
\end{equation}
one obtains the expression of the boundary action upto quadratic in 
 $b_{10}$, $b_{20}$,$b_{30}$. This expression is too long and so we are not writing
it explicitly. The R-charge conductivity can be obtained by
differentiating boundary action twice and are given by
\begin{equation}
\begin{split}
\lambda_{11} &=  
\dfrac{N_c^{(3/2)}T_0}{36\sqrt{2}}
\bigg[k_1^2(-1+k_2k_3) + k_1^2(k_2^2k-3 - 2(5+3 k_3) 
+ k_2(-6+k_3^2))\\
&-
4(2+2k_3+k_3^2+k_2^2(1+k_3)+k_2(2+2k_3+k_3^2)
-k_1(16 + 14 k_3 + 5 k_3^2 + k_2^2(5+4 k_3) + \\&2 k_2 (7+5 k_3 + 2 k_3^2)\bigg]
\dfrac{v}{2(1+k_1)(1+k_2)(1+k_3)},
\end{split}
\end{equation}
\begin{equation}
\begin{split}
\lambda_{12} &=  \dfrac{N_c^{(3/2)}T_0}{36\sqrt{2}}~ \dfrac{1}{4} 
\bigg[k_1(4+k_1+k_2+k_3)(-2-k_2-k_3+k_1(-1+k_2k_3)
(1+k_3))\bigg]\\&
\left(\dfrac{v}{(1+k_1)^2} + \dfrac{v}{(1+k_2)^2}\right),
\end{split}
\end{equation}

Other components of the conductivity matrix can be obtained from above two components
by cyclically permuting $k_1$, $k_2$ and $k_3$.
The diffusion coefficients can be obtained from the relation 
$D_{ij}=\sum\limits_{k=1}^3 \lambda_{ik}\chi^{-1}_{kj}$,
where $\chi$ is the specific heat. 
Since all the components of specific heat diverges
the diffusion coefficients turn out to be zero, as expected.

\section{Four dimensional black hole with three charges}

For four dimensional black hole with three charges we have $c=0,n=2,b=1$, $m=3$ and $l=1$. From $f(u)= (1-u)[1+(1+k_1+k_2+k_3) u +(1+k_1+k_2+k_3+ k_1 k_2+ k_2 k_3+k_1 k_3) u^{2})$ we get 
\begin{equation}
F(u)=1+(1+k_1+k_2+k_3) u +(1+k_1+k_2+k_3+ k_1 k_2+ k_2 k_3+k_1 k_3) u^{2}),
\end{equation}
which is quadratic. 
The solutions are given by
\begin{equation}
p_1(u) = b_{10}(1 + \frac{2k_1}{3}) + \dfrac{(1+k_1)}{3(1+k_2)(1+k_3)}
\Big[k_2(1+k_3)b_{20} + k_3(1+k_2)b_{30}\Big] ,
\end{equation} 
while other $p_i$'s can be obtained from above expression by permuting 
$(123)$ cyclically.

Using (\ref{generals}) $s_i(u)$'s is given by,
\begin{equation}
\begin{split}
s_1(u) &= h_1 u + (h_2 + h_3 u)\triangle(u) + ( h_4 + h_5 u) 
\log F(u),\\
s_2(u) &= l_1 u + (l_2 + l_3 u)\triangle(u) + ( l_4 + l_5 u) 
\log F(u),\\
s_3(u) &= m_1 u + (m_2 + m_3 u)\triangle(u) + ( m_4 + m_5 u) 
\log F(u),\\
\triangle(u) &= \arctan{\left[
\dfrac{
\sqrt{3 - k_1^2 - k_2^2 + 2 k_3 - k_3^2 + 2 k_2 ( 1 + k_3) + 2 k_1 (1 + k_2 + k_3)}
}{2+ (1+k_1+k_2+k_3)u}u\right]}.
\end{split}
\end{equation}
Here once again we have combined two logarithms as in (\ref{logcombine}).
The values of the parameters are as follows:
\begin{equation}
\begin{split}
h_3 &= \dfrac{\Big[k_3 - 2k_1(1+k_2)(1+k_3)\Big]h_2 + (1+k_1)k_2(1+k_3)l_2 + (1+k_1)k_2k_3 m_2}{3(1+k_2)(1+k_3)} v \\
h_4 &=\dfrac{3v}{2} b_{10}\\
h_5 &= \dfrac{2k_1(1+k_2)(1+k_3)b_{10} + (1+k_1)[k_2(1+k_3)b_{30} + k_3(1+k_2)b_{30}]}{2(1+k_2)(1+k_3)} v,
\end{split}
\end{equation}
\begin{equation}
\begin{split}
h_2 &=\bigg( 4(1+k_1)\Big[ 3+2k_1+2k_2+2k_3+2k_1k_2+2k_2k_3+2k_3k_1 \Big]\\
&\Big[ k_2(1+k_3)(1+k_1+k_2-k_3)b_{20} + k_3(1+k_2)(1+k_1-k_2+k_3)b_{30}\Big]\\
&-\Big[(1+k_2)(1+k_3)[4+k_1^4+k_2^4+k_3^4+4k_3+6k_3^2-2k_2^3(1+k_3) + 2k_1^3(7+3k_2+3k_3)+\\
&2k_2^2(-3-k_3+k_3^2)-2k_2(-3-3k_3+k_3^2+k_3^3)+ 2k_1^2(17+k_2^2+11k_3+k_3^2+11k_2+3k_2k_3)\\
&-2k_1(-15+k_2^3-11k_3+k_3^2+k_3^3+k_2^2+k_2^2k_3-11k_2-4k_2k_3+k_2k_3^2)\Big] \bigg) v \\
&\times\bigg((1+k_2)(1+k_3)\Big[k_1^3+k_2^3-k_2^2(1+k_3)+ (k_3 - 3)(1+k_3)^2-k_1^2(1+k_2+k_3)\\&-k_2(5+6k_3+k_3^2)
-k_1(5+k_2^2+6k_3+k_3^2+6k_2+6k_2k_3)\Big]\\
&\times \sqrt{(3-k_1^2-k_2^2-k_3^2+2k_1+2k_2+2k_3+2k_1k_2+2k_2k_3+2k_3k_1)}\bigg)^{-1},
\end{split}
\end{equation}
where we have used $v= T_0/(2T)$.
The components of R-charge conductivity turn out to be
\begin{equation}
\begin{split}
\lambda_{11} &= 2\bigg( 4 k_1^4\Big[2+k_2+k_3\Big]+k_1^3\Big[53+5k_2^2+ 43 k_3 + 5 k_3^2 + k_2(43 + 23 k_3)\Big]\\
&+ k_1^2 \Big[ 108 + k_2^3 +130 k_3 + 43 k_3^2 + k_3^3 + k_2^2(43+26 k_3) + 2 k_2(65 + 60 k_3  + 13 k_3^2)\Big]\\
&+9\Big[3+ 6 k_3 + 4 k_3^2 + k_3^3 + k_2^3 (1+k_3) + k_2^2(4 + 6 k_3 + 3 k_3^2) + k_2(6+10 k_3 + 6 k_3^2 + k_3^3)\Big]\\
&k_1\Big[90 + 144 k_3 + 74 k_3^2 + 11 k_3^3 + k_2^3(11+10 k_3) \\&
+ k_2^2(74 + 82k_3 + 29 k_3^2) + k_2(144 + 187 k_3 + 82 k_3^2 + 10 k_3^3)\Big]\bigg)\\
\times & \bigg(\dfrac{v}{9(1+k_1)(1+k_2)(1+k_3)(1+k_1+k_2+k_3)}\bigg),
\end{split}
\end{equation}
\begin{equation}
\begin{split}
\lambda_{12} &= 
-\dfrac{\Big[ k_2 + k_1^2 k_2 + k_1(1+4k_2 + k_2^2)\Big] v}{9(1+k_1)^2(1+k_2)^2(1+k_3)(1+k_1+k_2+k_3)} 
\bigg( 2 k_2^3\Big[2+k_3\Big] + 
2 k_1^3\Big[2+k_2+k_3\Big] +
\\
& k_2^3\Big[22+14k_3 + k_3^2\Big] - 2\Big[-9 -9k_3 + k_3^2 + k_3^3\Big] - k_2\Big[-36 - 26 k_3 - 4k_3^2 + k_3^3\Big] +\\ 
&
k_2^2 \Big[22+4k_2^2+14k_3+k_3^2+k_2(20+7k_3)\Big] -
2\Big[-9 -9 k_3 + k_3^2 + k_3^3\Big] - \\
&k_2 \Big[-36 - 26 k_3 + 4k_3^2 + k_3^3\Big] + 
k_1^2\Big[22 + 4k_2^2 + 14 k_3 + k_3^2 + k_2(20+7k_3)\Big]\\
&+ k_1\Big[36 + 2 k_2^3 + 26 k_3 - 4k_3^2 -k_3^3+ k_2^2(20+7k_3) + k_2(50 + 24k_3 - 8k_3^2)\Big] \bigg)
\end{split}
\end{equation}
while other components can be obtained by cyclically 
permuting $k_1$, $k_2$ and $k_3$.

\section{Four dimensional black hole with four charges}

For four dimensional black hole with four charges we have $c=0, n=2, b=1$, $m=4$ and $l=1$. $f(u)$ is given by
\begin{equation}
\begin{split}
f&=(1-u)[1+(1+k_1+k_2+k_3+k_4) u +
\\&
(1+k_1+k_2+k_3+k_4+ k_1 k_2+ k_1 k_3+k_1 k_4+k_2 k_3+k_2 k_4+ k_3 k_4) u^{2}-k_1k_2k_3k_4 u^{3}],
\end{split}
\end{equation}
and so we have
\begin{equation}
\begin{split}
F(u)&=1+(1+k_1+k_2+k_3+k_4) u +
\\&
(1+k_1+k_2+k_3+k_4+ k_1 k_2+ k_1 k_3+k_1 k_4+k_2 k_3+k_2 k_4+ k_3 k_4) u^{2}-k_1k_2k_3k_4 u^{3}.
\end{split}
\end{equation}
which is a cubic polynomial.

From (\ref{generalp}) we can write
\begin{equation}
p_i(u) = b_{i0} + c_{i0} u,
\end{equation}
where $b_{i0}$ are constant parameters,
$c_{10}$ is given by
\begin{equation}
c_{10}=\dfrac{2k_1}{3}b_{10} - \dfrac{1+k_1}{3}[\dfrac{k_2}{1+k_2} b_{20}
-\dfrac{k_3}{1+k_3} b_{30} - \dfrac{k_4}{1+k_4} b_{40}]
\end{equation}
and other $c_{i0}$ can be obtained by permuting $(1234)$ in the above expression.
From (\ref{generals}) we obtain
\begin{equation}
\begin{split}
s_1(u) &= h_1 u + (h_2+h_3u)\log(1-u/u_1)
+ (h_4 + h_5 u)\log (1-u/u_2)\\ &+ (h_6 + h_7 u) \log(1-u/u_3)
\end{split}
\end{equation}
Since $F(u)$ is a polynomial of degree 3 the zeroes of $F(u)$, $u_i, i=1,2,3$ have complicated expressions.
 Because of the complicated, which in turn makes explicit expressions of parameters $h_1$ to $h_7$  too long to include here.
Once again proceeding in a similar way we obtain the
R-charge conductivities in terms of $k_i$'s.

\section{Seven dimensional black hole with two charges}
In the case of seven dimensional black hole with two charges we have $c = -1$, $n=3$, $b=4$ and $m=3$.
\begin{equation}
f=(1-u)[1+ u +(1+ k_1+ k_2+ k_1 k_2) u^{2}-k_1k_2 u^{3}].
\end{equation}
So $F(u)$ is given by
\begin{equation}
F(u)=1+ u +(1+ k_1+ k_2+ k_1 k_2) u^{2}-k_1k_2 u^{3},
\end{equation}
which is, once again, cubic.

Since $l=2$ from (\ref{generalp}) we get
\begin{equation}
p_i(u) = b_{i0} + c_{i0} u^{2}.
\end{equation}
From (\ref{generals}) we get 
\begin{equation}
\begin{split}
s_1(u) &= h_1 u +h_2 u^{2} + (h_3+h_4 u^{2})\log (1-u/u_1)
+ (g_5 + g_6 u^{2})[\log (1-u/u_2)\\ &+ (g_7 + g_8 u^{2}) \log (1- u/u_3).
\end{split}
\end{equation}
$F(u)$ being of order 3, expressions of zeroes $u_i$'s are more involved which makes the expressions of  $h_1$ to $h_8$ too long and complicated to write down explicitly. We have analyzed this case and obtain similar conclusion that as critical surfaces are approached diffusion coefficients vanish.

\end{document}